\documentclass[prd,showpacs,floatfix,amsmath,amssymb]{revtex4}
\usepackage{bm}
\usepackage{hyperref}
\usepackage{graphicx}
\newcommand{\ud}{\mathrm{d}} 

\begin{document}
\title{Neutrino Processes in Strong Magnetic Fields 
and Implications for Supernova Dynamics}
\author{Huaiyu Duan}
\email{hyduan@physics.umn.edu}
\author{Yong-Zhong Qian}
\email{qian@physics.umn.edu}
\affiliation{
School of Physics and Astronomy, University of Minnesota,
Minneapolis, MN 55455}
\date{\today}
 
\begin{abstract}
The processes $\nu_e+n\rightleftharpoons p+e^-$ and 
$\bar\nu_e+p\rightleftharpoons n+e^+$ provide the dominant mechanisms
for heating and cooling the material between the protoneutron star
and the stalled shock in a core-collapse supernova. Observations
suggest that some neutron stars are born with magnetic fields of
at least $\sim 10^{15}$ G while theoretical considerations give an
upper limit of $\sim 10^{18}$ G for the protoneutron star magnetic 
fields. We calculate the rates for the above neutrino processes in 
strong magnetic fields of $\sim 10^{16}$ G. 
We find that the main effect of such magnetic fields is to change the
equations of state through the phase space of $e^-$ and $e^+$, which
differs from the classical case due to quantization of the motion of 
$e^-$ and $e^+$ perpendicular to the magnetic field. As a result,
the cooling rate can be greatly reduced by magnetic fields of 
$\sim 10^{16}$ G for typical conditions below the stalled shock
and a nonuniform protoneutron star magnetic field (e.g., a
dipole field) can introduce a large angular dependence of the
cooling rate. In addition, strong magnetic fields always lead to an
angle-dependent heating rate by polarizing the spin of $n$ and $p$. 
The implications of our results for the neutrino-driven supernova 
mechanism are discussed.
\end{abstract}
\maketitle

\section{Introduction}

In this paper we study neutrino processes in strong magnetic fields
and their implications for supernova dynamics. Although the detailed 
mechanism by which massive stars produce supernova explosions is 
still elusive (see Ref.~\cite{Bethe:1990mw} for a review), 
intense research in the past few decades 
has led to the following prevalent paradigm.
At the exhaustion of nuclear fuels, the Fe core of a massive star 
collapses. When nuclear density is reached, the inner core bounces
and a shock is launched. As the shock propagates outward,
it loses energy by dissociating the Fe nuclei falling through it. 
Eventually, the shock is stalled before exiting the outer core.
Meanwhile, the inner core is settling into a protoneutron star
by emitting $\nu_e$, $\bar\nu_e$, $\nu_\mu$, $\bar\nu_\mu$, $\nu_\tau$, 
and $\bar\nu_\tau$. These neutrinos can exchange energy with the
material below the stalled shock. The dominant energy-exchange 
processes are 
\begin{eqnarray}
\nu_e+n&\rightleftharpoons&p+e^-,\label{eq:nun}\\ 
\bar\nu_e+p&\rightleftharpoons&n+e^+.\label{eq:nup}
\end{eqnarray}
The forward processes in Eqs.~\eqref{eq:nun} and \eqref{eq:nup} heat the 
material through absorption of $\nu_e$ and $\bar{\nu}_e$
while the reverse processes cool
the material through capture of $e^-$ and $e^+$. The competition between 
heating and cooling of the material by these processes is expected
to result in net energy gain for the shock, which then propagates 
outward again to make a supernova explosion. This is the
neutrino-driven supernova mechanism \cite{Bethe:1985}.

Unfortunately, the current consensus is that the neutrino-driven 
supernova mechanism does not work in spherically symmetric models 
\cite{Rampp:2000,Liebendorfer:2001}. 
One group has shown that this mechanism works in 
three-dimensional models where spherical symmetry is broken by 
convection \cite{Fryer:2002}. However, this success has not been confirmed 
by other groups yet. On the other hand, magnetic fields may be 
generated during the formation of protoneutron stars and in turn affect
supernova dynamics. Magnetic fields of 
$\sim 10^{12}$ G are commonly inferred for pulsars. Observations 
also suggest that a number of neutron stars, the so-called magnetars, 
have magnetic fields of $\sim 10^{15}$ G 
(see e.g., Refs.~\cite{Kouveliotou:1999,Gotthelf:1999,Ibrahim:2003}). 
A theoretical
upper limit of $\sim 10^{18}$ G may be estimated for the magnetic fields 
of new-born neutron stars \cite{Lai:2000at}. While 
strong magnetic fields may induce supernova explosions directly 
through dynamic effects such as jet production \cite{Khokhlov:1999}, and
therefore, render the neutrino-driven mechanism irrelevant, the details
of this magnetohydrodynamic mechanism have not been worked out or
understood yet. In this paper we address the effects of strong 
magnetic fields on supernova dynamics still in the context of the
neutrino-driven mechanism. In particular, we focus on how such fields 
affect the microscopic processes of heating and cooling the material
below the stalled shock.

We describe the neutrino-driven supernova mechanism without magnetic
fields in some detail in Sec.~\ref{sec:supernova-mechanism}. 
The rates of the processes in
Eqs.~\eqref{eq:nun} and \eqref{eq:nup} in strong
magnetic fields are calculated in Sec.~\ref{sec:rates-in-fields}. 
The implications of
these rates for supernova dynamics are discussed in 
Sec.~\ref{sec:implications} and
conclusions given in Sec.~\ref{sec:conclusions}.

\section{\label{sec:supernova-mechanism}
The Neutrino-Driven Supernova Mechanism}

\begin{figure}
\includegraphics*[width=2.5in, keepaspectratio]{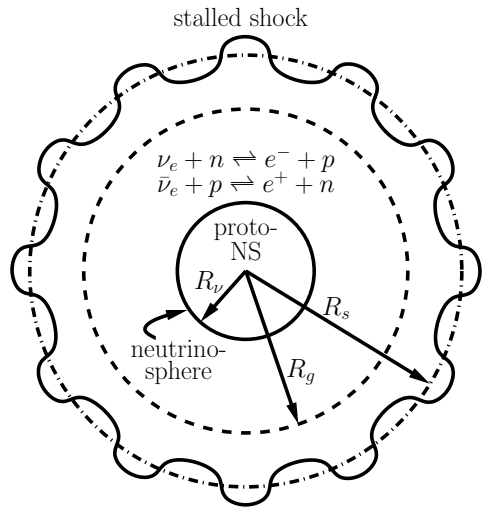}
\caption{\label{fig:1}A sketch of the region of interest. 
The neutrinosphere of radius
$R_\nu=50$ km effectively defines the surface of the protoneutron star.
The stalled shock is at an average radius $R_s=200$ km. At the gain
radius $R_g$, the rate for heating by absorption of $\nu_e$ and
$\bar\nu_e$ equals that for cooling by capture of $e^-$ and $e^+$.
Heating dominates cooling above $R_g$.}
\end{figure}

In this section we give a more quantitative description of the 
neutrino-driven supernova mechanism in the absence of magnetic fields.
As this mechanism has not been fully established yet, we will use
parameters typical of current models to illustrate the essence of
these models rather than focus on the numerical details of a specific
model. We are interested in times of $\lesssim 0.5$ s after the core
collapse, which correspond to the critical period for the neutrino-driven 
supernova mechanism.
The region of interest is above the protoneutron star but below
the stalled shock as illustrated in Fig.~\ref{fig:1}. We consider that 
neutrinos are emitted from a neutrinosphere of radius $R_\nu=50$ km 
that effectively defines the surface of the protoneutron star
(see e.g., Fig.~11.1 in Ref.~\cite{Raffelt:1996}).
The stalled shock is taken to be at an average radius $R_s=200$ km.
The material in the region of interest has typical entropies of 
$\sim 10$ (in units of Boltzmann constant per nucleon) 
and typical temperatures of several MeV. For these conditions, 
the material can be characterized as a gas of $n$, $p$, $e^-$, $e^+$,
and $\gamma$ (photons). The predominant cooling processes are the 
reverse reactions in Eqs.~\eqref{eq:nun} and \eqref{eq:nup} while
the predominant heating processes are the corresponding forward 
reactions [note that similar charged-current processes involving
$\nu_{\mu(\tau)}$ and $\bar\nu_{\mu(\tau)}$ are energetically 
forbidden for the neutrino energies and material conditions available
in supernovae]. Cooling dominates heating near the neutrinosphere. 
However, the cooling rate decreases 
much more steeply with increasing radius than the heating rate. 
These two rates becomes equal at the gain radius $R_g$, above which
heating dominates. Thus, the heating and cooling rates 
are crucial to a quantitative discussion of the neutrino-driven 
supernova mechanism. These rates are calculated below.

We start with a description of neutrino emission by the protoneutron 
star. The $\nu_e$ luminosity 
$L_{\nu_e}$ is taken to be the same as the $\bar\nu_e$ luminosity 
$L_{\bar\nu_e}$. We assume $L_{\nu_e}=L_{\bar\nu_e}=4\times 10^{52}$ 
erg s$^{-1}$ during the epoch relevant for shock revival by neutrino
heating. As $\nu_e$
and $\bar\nu_e$ are roughly in thermal equilibrium with the matter
at the neutrinosphere, their luminosities approximately correspond to
those of black-body radiation for a Fermi-Dirac neutrino energy 
distribution with zero chemical potential:
\begin{widetext}
\begin{equation}
L_{\nu_e}=L_{\bar\nu_e}\sim\frac{7\pi^3}{240}
[T(R_\nu)]^4R_\nu^2\sim 
4\times 10^{52}\left[\frac{T(R_\nu)}{4\ \mathrm{MeV}}\right]^4
\left(\frac{R_\nu}{50\ \mathrm{km}}\right)^2\ \mathrm{erg\ s}^{-1}.
\label{eq:lnu}
\end{equation}
\end{widetext}
In the above equation, $T(R_\nu)$ is the temperature at the 
neutrinosphere. Throughout this
paper, we adopt units where the Planck constant $\hbar$, the speed
of light $c$, and the Boltzmann constant $k$ are set to unity.

Due to the difference in the interaction of $\bar\nu_e$ and $\nu_e$
with the protoneutron 
star matter, their emission is more complicated
than implied by the crude estimate in Eq.~(\ref{eq:lnu}). As there
are fewer protons to absorb $\bar\nu_e$ than neutrons to absorb
$\nu_e$, $\bar\nu_e$ decouple from the protoneutron 
star matter at 
higher temperature and density than $\nu_e$. This results in a 
higher average energy for $\bar\nu_e$ (see below). However, due to 
the steep temperature and density gradients near the protoneutron 
star surface, the radii for $\bar\nu_e$ and $\nu_e$ decoupling
are essentially the same. Detailed neutrino transport calculations 
show that $\bar\nu_e$ and $\nu_e$ can be considered as having the
same luminosity and neutrinosphere but significantly different
average energies (see e.g., Ref.~\cite{Janka:1995}). The normalized 
$\nu_e$ and $\bar\nu_e$ energy distributions can be described 
by functions of the form
\begin{equation}
f_{\nu}(E_\nu)=\frac{1}{F_2(\eta_\nu)T_\nu^3}
\frac{E_\nu^2}{\exp[(E_\nu/T_\nu)-\eta_\nu]+1},
\end{equation}
where the subscript $\nu$ refers to $\nu_e$ or $\bar\nu_e$, $E_\nu$
is the neutrino energy, $T_\nu$ and $\eta_\nu$ are two positive 
parameters, and $F_2(\eta_\nu)$ is a specific case of the general 
Fermi integral $F_n(\eta)$ defined as
\begin{equation}
F_n(\eta)\equiv\int_0^\infty\frac{x^n}{\exp(x-\eta)+1} \, \ud x.
\end{equation}
The parameter $T_\nu$ is related to the average neutrino energy 
$\langle E_\nu\rangle$ as
\begin{equation}
\langle E_\nu\rangle=\frac{F_3(\eta_\nu)}{F_2(\eta_\nu)}T_\nu.
\end{equation}
We take $\eta_{\nu_e}=\eta_{\bar\nu_e}=3$, 
$\langle E_{\nu_e}\rangle=11$ MeV, and 
$\langle E_{\bar\nu_e}\rangle=16$ MeV. For these parameters, 
$T_{\nu_e}=2.75$ MeV and $T_{\bar\nu_e}=4$ MeV, which are close to
$T(R_\nu)$ estimated from Eq.~\eqref{eq:lnu}.

\begin{figure}
\includegraphics*[width=3.25in, keepaspectratio]{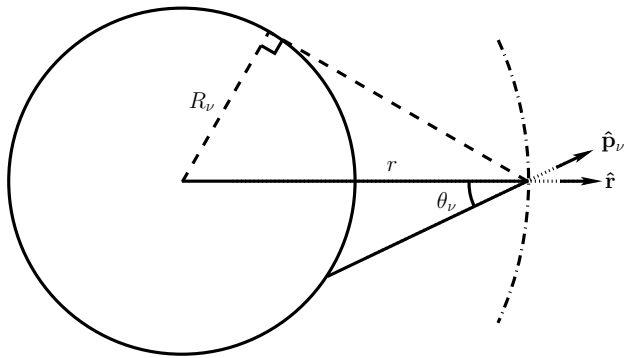}
\caption{\label{fig:2}A sketch of the geometry for calculating the differential
neutrino number density $\ud^2n_\nu/\ud E_\nu \ud\Omega_\nu$ at radius $r$.
For a specific radial direction $\bf{\hat r}$, the differential solid
angle $\ud\Omega_\nu$ is defined by the polar angle
$\theta_\nu$ between $\bf{\hat r}$ and the direction of the neutrino
momentum $\bf{\hat p}_\nu$ (the corresponding azimuthal angle $\phi_\nu$
is not shown). Only neutrinos emitted with
$\sqrt{1-(R_\nu/r)^2}\leq\cos\theta_\nu\leq 1$ can contribute to
$\ud^2n_\nu/\ud E_\nu \ud\Omega_\nu$.}
\end{figure}

To calculate the rate of heating by neutrino absorption processes,
we need the differential neutrino number density 
per unit energy interval and per unit solid angle 
$\ud^2 n_\nu / \ud E_\nu \ud\Omega_\nu$ at radius $r>R_\nu$ as measured
from the center of the protoneutron star.
For a specific radial direction $\mathbf{\hat r}$, the differential solid
angle $\ud\Omega_\nu$ is defined by the polar angle $\theta_\nu$
between $\mathbf{\hat r}$ and the direction of the neutrino momentum
$\mathbf{\hat p}_\nu$ (see Fig.~\ref{fig:2})
and by the corresponding azimuthal angle $\phi_\nu$. 
We assume that only neutrinos emitted with
$\sqrt{1-(R_\nu/r)^2}\leq\cos\theta_\nu\leq 1$ can contribute
to $\ud^2n_\nu/\ud E_\nu \ud\Omega_\nu$ (see Fig.~\ref{fig:2}). Thus,
\begin{equation}
\frac{\ud^2n_\nu}{\ud E_\nu \ud\Omega_\nu}=\left\{\begin{array}{ll}
L_\nu f_\nu(E_\nu)/(4\pi^2R_\nu^2\langle E_\nu\rangle),&
\text{for }\sqrt{1-(R_\nu/r)^2}\leq\cos\theta_\nu\leq 1,\\
0,&\text{otherwise}.\end{array}\right.
\end{equation}
The corresponding neutrino occupation number $\tilde f_\nu(E_\nu)$ is
\begin{equation}
\tilde f_\nu(E_\nu)=\frac{(2\pi)^3}{E_\nu^2}
\frac{\ud^2n_\nu}{\ud E_\nu \ud\Omega_\nu}.
\label{eq:nuoccu}
\end{equation}
Note that 
\begin{equation}
L_\nu=4\pi R_\nu^2
\int_0^{2\pi}\ud\phi_\nu\int_0^1\cos\theta_\nu \, \ud\!\cos\theta_\nu
\int_0^\infty E_\nu\frac{\ud^2n_\nu}{\ud E_\nu \ud\Omega_\nu} \, \ud E_\nu,
\end{equation}
where $\ud^2n_\nu/\ud E_\nu \ud\Omega_\nu$ is evaluated at $r=R_\nu$.

To zeroth order in $m_N^{-1}$ where $m_N$ is the nucleon mass,
the cross sections for the forward processes in Eqs.~\eqref{eq:nun}
and \eqref{eq:nup} are
\begin{equation}
\sigma_{\nu N}^{(0)}=\frac{G_F^2\cos^2\theta_C}{\pi}(f^2+3g^2)p_e E_e,
\label{eq:sigh}
\end{equation}
where the subscript $\nu N$ refers to $\nu_e$ absorption on $n$ or 
$\bar\nu_e$ absorption on $p$,
$G_F=(292.8\ \mathrm{GeV})^{-2}$ is the Fermi constant, $\theta_C$ 
is the Cabbibo angle ($\cos^2\theta_C=0.95$), $f=1$ and $g=1.26$ are
the vector and axial-vector coupling coefficients, respectivley,
of the weak interaction, and $p_e$ and $E_e=\sqrt{p_e^2+m_e^2}$ 
are the momentum and energy, respectively, of the electron or 
positron in the final state ($m_e$ is the electron mass). 
To the same order in $m_N^{-1}$, conservation of
momentum and energy gives
\begin{equation}
E_e=E_\nu\pm\Delta, 
\end{equation}
where $\Delta=1.293$ MeV is the neutron-proton mass difference,
the plus sign is for absorption of $\nu_e$, and the minus sign is
for absorption of $\bar\nu_e$. Note that there is a threshold energy 
of $\Delta+m_e$ for $\bar\nu_e$ absorption on $p$.

\begin{figure}
\includegraphics*[width=3.25in, keepaspectratio]{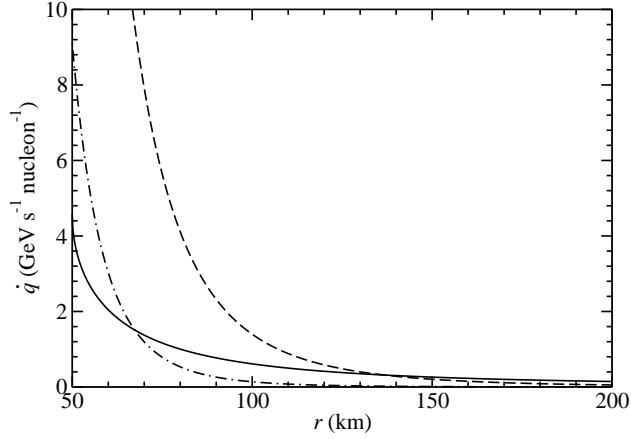}
\caption{\label{fig:3}The heating rate per nucleon $\dot q_h^{(0)}$ 
(solid curve)
and the cooling rate per nucleon $\dot q_c^{(0)}$ [dashed ($S=10$) and
dot-dashed ($S=20$) curves] as functions of radius $r$. The gain radius
is at $R_g=137$ and 68 km for entropies of $S=10$ and 20, respectively.}
\end{figure}

At radius $r>R_\nu$, the heating rate per nucleon $\dot q_h^{(0)}$ 
due to absorption of $\nu_e$ and $\bar\nu_e$ is
\begin{eqnarray}
\dot q_h^{(0)}&=&Y_n\int \ud\Omega_{\nu_e}\int_0^\infty(E_{\nu_e}+\Delta)
\sigma_{\nu_e n}^{(0)}\frac{\ud^2n_{\nu_e}}{\ud E_{\nu_e}\ud\Omega_{\nu_e}}
\,\ud E_{\nu_e}\nonumber\\
&+&Y_p\int \ud\Omega_{\bar\nu_e}\int_{\Delta+m_e}^\infty
(E_{\bar\nu_e}-\Delta)\sigma_{\bar\nu_e p}^{(0)}
\frac{\ud^2n_{\bar\nu_e}}{\ud E_{\bar\nu_e}\ud\Omega_{\bar\nu_e}}
\,\ud E_{\bar\nu_e}\nonumber\\
&=&(3.83\times 10^3Y_n+5.32\times 10^3Y_p)
\left[1-\sqrt{1-(R_\nu/r)^2}\right]
\ \mathrm{MeV\ s}^{-1}\ \mathrm{nucleon}^{-1},\label{eq:qh}
\end{eqnarray}
where the numerical coefficients correspond to the parameters
$L_{\nu_e}$, $L_{\bar\nu_e}$, $T_{\nu_e}$, $T_{\bar\nu_e}$,
$\eta_{\nu_e}$, and $\eta_{\bar\nu_e}$ adopted above, and
$Y_n$ and $Y_p$ are the neutron and proton number fractions,
respectively. As the material is neutral, $Y_n=1-Y_e$ and
$Y_p=Y_e$, where $Y_e$ is the net electron number per nucleon.
The radial dependence in Eq.~\eqref{eq:qh} comes from the integration
over $\ud\Omega_\nu$ and accounts for the geometric dilution of the
neutrino number density. Note that for $r\gg R_\nu$, 
$\dot q_h^{(0)}\propto r^{-2}$. Taking $Y_e=0.5$, we calculate
$\dot q_h^{(0)}$ from Eq.~\eqref{eq:qh} and show the result as the
solid curve in Fig.~\ref{fig:3}.

Next we calculate the cooling rate due to the reverse 
processes in Eqs.~\eqref{eq:nun} and \eqref{eq:nup}. It is
convenient to define a volume reaction rate, which gives the rate
of e.g., $e^+$ capture per neutron when multiplied by the $e^+$
number density $n_{e^+}$. In the absence of magnetic 
field, the volume reaction rate $\Gamma_{eN}^{(0)}$
is simply $v\sigma_{eN}^{(0)}$,
where $v$ is the relative velocity between the reacting particles
and $\sigma_{eN}^{(0)}$ is the cross section for $e^+$ capture on $n$ or 
$e^-$ capture on $p$. To zeroth order in $m_N^{-1}$, the volume reaction 
rates for these processes are
\begin{equation}
\Gamma_{eN}^{(0)}=\frac{G_F^2\cos^2\theta_C}{2\pi}(f^2+3g^2)E_\nu^2.
\label{eq:sigc}
\end{equation}
In Eq.~\eqref{eq:sigc}, 
\begin{equation}
E_\nu=E_e\pm\Delta,
\end{equation}
where the plus sign is for $e^+$ capture on $n$ and the minus sign is for
$e^-$ capture on $p$. The cooling rate per nucleon $\dot q_c^{(0)}$ is then
\begin{equation}
\dot q_c^{(0)}=\frac{Y_n}{\pi^2}\int_0^\infty
\frac{E_{e^+}\Gamma_{e^+n}^{(0)}p_{e^+}^2}{\exp[(E_{e^+}/T)+\eta_e]+1}
\,\ud p_{e^+}+\frac{Y_p}{\pi^2}\int_{\sqrt{\Delta^2-m_e^2}}^\infty 
\frac{E_{e^-}\Gamma_{e^-p}^{(0)}p_{e^-}^2}{\exp[(E_{e^-}/T)-\eta_e]+1}
\,\ud p_{e^-},
\label{eq:qc}
\end{equation}
where $T$ is the temperature, $\eta_e$ is the electron degeneracy
parameter, and $\sqrt{\Delta^2-m_e^2}$ is the threshold momentum for
$e^-$ capture on $p$.

To evaluate $\dot q_c^{(0)}$ we need $T$ and $\eta_e$. We take
\begin{equation}
T(r)=T(R_\nu)\frac{R_\nu}{r}=
4\left(\frac{50\ \mathrm{km}}{r}\right)\ \mathrm{MeV}
\label{eq:tr}
\end{equation}
for $R_\nu\leq r\leq R_s$ (see e.g., Ref.~\cite{Janka:2001}). 
We assume that the material in the region
of interest can be characterized by a typical electron fraction $Y_e$ 
and a typical entropy per nucleon $S$. We then obtain $\eta_e$
together with the matter density $\rho$ from the equations of state:
\begin{eqnarray}
\frac{\rho Y_e}{m_N}&=&n_{e^-}-n_{e^+},\label{eq:eosye}\\
S&=&S_N+S_\gamma+S_{e^-}+S_{e^+},
\label{eq:eoss}
\end{eqnarray}
where $S_N$, $S_\gamma$, $S_{e^-}$, and $S_{e^+}$ are the contributions 
to $S$ from nucleons, photons, electrons, and 
positrons, respectively. The expressions for $n_{e^-}$, $n_{e^+}$,
$S_N$, $S_\gamma$, $S_{e^-}$, and $S_{e^+}$ are given in Appendix 
\ref{sec:eos}.
We note that for extremely relativistic $e^-$ and $e^+$,
\begin{eqnarray}
n_{e^-}-n_{e^+}&=&\frac{\eta_e}{3}
\left(1+\frac{\eta_e}{\pi^2}\right)T^3,\label{eq:ne}\\
S_{e^-}+S_{e^+}&=&\frac{7\pi^2}{45}
\left(1+\frac{15\eta_e^2}{7\pi^2}\right)\left(\frac{m_N}{\rho}\right)T^3.
\label{eq:se}
\end{eqnarray}
While we always use the expressions in Appendix \ref{sec:eos} to calculate 
the results presented in this paper, we find that Eqs. \eqref{eq:ne} and 
\eqref{eq:se} are excellent approximations to the corresponding general
expressions (without magnetic fields) for the conditions in the region 
of interest.

\begin{figure*}
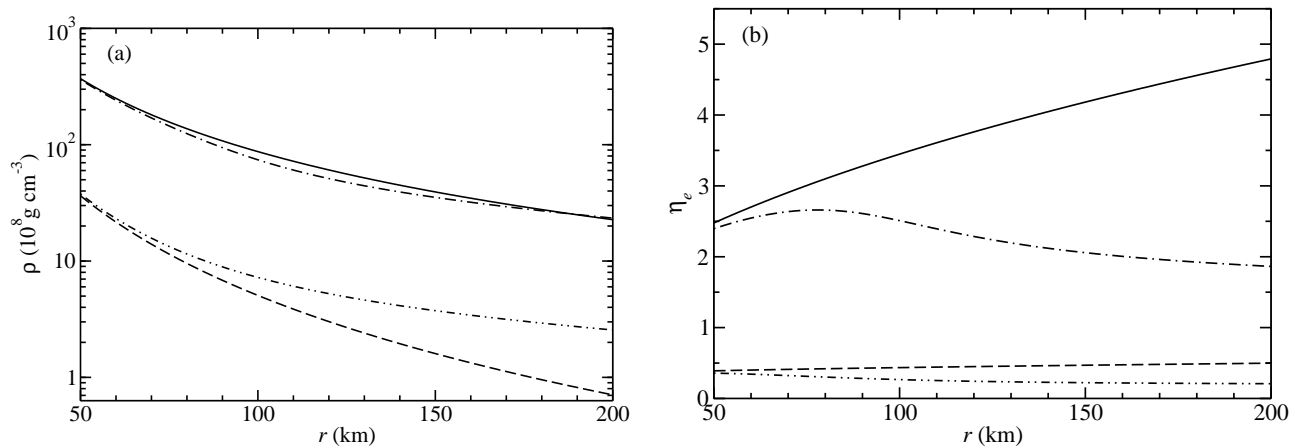

\begin{center}
$\begin{array}{@{}c@{\hspace{.2in}}c@{}}
\includegraphics*[width=3.25in, keepaspectratio]{fig4a.eps} &
\includegraphics*[width=3.25in, keepaspectratio]{fig4b.eps}
\end{array}$
\end{center}
\caption{\label{fig:4}The matter density $\rho$ (a) and the electron 
degeneracy
parameter
$\eta_e$ (b) as functions of radius $r$. The solid ($S=10$) and dashed
($S=20$) curves are for the case of no magnetic field. The dot-dashed
($S=10$) and dot-dot-dashed ($S=20$) curves are for the case of a uniform
magnetic field of $B=10^{16}$ G. The magnetic field greatly reduces
$\eta_e$ for an entropy of $S=10$. For $S=20$, $\eta_e$ is very small
even for the case of no magnetic field.}
\end{figure*}

Numerical models \cite{Rampp:2000,Liebendorfer:2001} show that $Y_e$ and $S$
tend to rise sharply over a short distance above the neutrinosphere
and then stay approximately constant. As cooling always dominates
near the neutrinosphere, the gain radius lies in the region where
$Y_e$ and $S$ can be taken as constant. Thus, we assume no radial
dependence for $Y_e$ and $S$ in determining the gain radius.
Taking $Y_e=0.5$ and $S=10$ and 20, we calculate $\rho$ and
$\eta_e$ as functions of $r$ from Eq.~\eqref{eq:tr} and the equations of
state [see Eqs.~\eqref{eq:eosye} and \eqref{eq:eoss} and Appendix 
\ref{sec:eos}]. 
The results are shown as the solid ($S=10$) and
dashed ($S=20$) curves in Fig.~\ref{fig:4}a for $\rho$ and 
Fig.~\ref{fig:4}b for $\eta_e$.
We also calculate the 
corresponding $\dot q_c^{(0)}$ from Eq.~\eqref{eq:qc} and show the results 
as the dashed ($S=10$) and dot-dashed ($S=20$) curves in
Fig.~\ref{fig:3} along with the result for $\dot q_h^{(0)}$ (solid curve). 
It can be seen that the 
gain radius is at $R_g=137$ and 68 km for
$S=10$ and 20, respectively. As the shock is at radius $R_s=200$ km,
there is a large region for net heating below the shock in both
cases. Note that the location of the gain radius is rather sensitive
to $S$. A gain radius below the shock ($R_g<R_s$) exists only for 
$S\gtrsim 10$. We have also done calculations for different values of
$Y_e$ and found that the effects of strong magnetic fields to be
discussed are qualitatively the same for $Y_e\gtrsim 0.3$. For
clarity of presentation, we focus on the results for $Y_e=0.5$.

In the above calculation of $\dot q_h^{(0)}$ and $\dot q_c^{(0)}$,
we have ignored Pauli blocking of the final states by the $e^-$, $e^+$,
$\nu_e$, and $\bar\nu_e$ (see e.g., Ref.~\cite{Gvozdev:2002})
in the region above the neutrinosphere. In the case of $\dot q_h^{(0)}$,
the $e^-$ and $e^+$ produced by $\nu_e$ and $\bar\nu_e$ absorption
have typical energies of $\sim 20$ MeV that are much higher than the
average energies of $e^-$ and $e^+$ in the gas. So Pauli blocking is
unimportant for calculating $\dot q_c^{(0)}$, especially in the region 
near and above the gain radius.
In the case of $\dot q_c^{(0)}$, the neutrino occupation number 
$\tilde f_\nu(E_\nu)$ can be calculated from Eq.~(\ref{eq:nuoccu}). 
For the adopted parameters, we find 
$\tilde f_{\nu_e}(E_{\nu_e})\leq 0.37$ and
$\tilde f_{\bar\nu_e}(E_{\bar\nu_e})\leq 0.083$ for
$\sqrt{1-(R_\nu/r)^2}\leq\cos\theta_\nu\leq 1$ and 
$\tilde f_{\nu_e}(E_{\nu_e})=\tilde f_{\bar\nu_e}(E_{\bar\nu_e})=0$
otherwise. As the range of $\cos\theta_\nu$ for finite 
$\tilde f_\nu(E_\nu)$ diminishes with increasing $r$,
Pauli blocking in this case is also insignificant in the region near 
and above the gain radius. Thus, ignoring Pauli blocking in the
calculation of $\dot q_h^{(0)}$ and $\dot q_c^{(0)}$ has little
effect on our discussion of the gain radius above. For the same
reasons, we will ignore Pauli blocking in the calculation of
the heating and cooling rates in strong magnetic fields.
This approximation will not affect the comparison of the gain 
radii for the cases without and with strong magnetic fields.

\section{\label{sec:rates-in-fields}
Heating and Cooling Rates in Magnetic Fields}

In this section we calculate the rates of heating and cooling by the
processes in Eqs.~\eqref{eq:nun} and \eqref{eq:nup} in strong
magnetic fields. We consider a uniform magnetic field of constant
strength $B$ in the $z$-direction. Observations indicate that
neutron stars may have magnetic fields up to $\sim 10^{15}$ G long after
their birth in supernovae.
This suggests that magnetic fields of at least $\sim 10^{15}$ G
can be generated during the formation of some neutron stars. An upper
limit of $\sim 10^{18}$ G for protoneutron star magnetic fields
can be estimated by equating the magnetic 
energy to the gravitational binding energy of a neutron star
\cite{Lai:2000at}. In this paper we consider protoneutron star magnetic fields 
of $B\sim 10^{16}$ G.

An obvious effect of the magnetic field is polarization of the spin of
a nonrelativistic nucleon due to the interaction Hamiltonian 
\begin{equation}
-\bm{\mu}\cdot \mathbf{B}=-\mu\sigma_zB.
\label{eq:mub}
\end{equation}
In Eq.~\eqref{eq:mub}, $\bm{\mu}=\mu\bm{\sigma}$ is the nucleon magnetic
moment, where $\mu=2.79\mu_N$ for $p$ and $\mu=-1.91\mu_N$ for $n$
with $\mu_N=e/(2m_p)$ being the nuclear magneton, and $\bm{\sigma}$ 
refers to the Pauli spin matrices. For a nucleon gas of temperature $T$,
the net polarization $\chi$ is
\begin{equation}
\chi=\frac{\exp(\mu B/T)-\exp(-\mu B/T)}{\exp(\mu B/T)+\exp(-\mu B/T)}.
\label{eq:chi}
\end{equation}
For $|\mu B/T|\ll 1$, Eq.~\eqref{eq:chi} reduces to
\begin{equation}
\chi=\frac{\mu B}{T}=3.15\times 10^{-2}\left(\frac{\mu}{\mu_N}\right)
\left(\frac{B}{10^{16}\ \mathrm{G}}\right)\left(\frac{\mathrm{MeV}}{T}\right).
\end{equation}

In addition, the motion of a proton in the $xy$-plane perpendicular
to the magnetic field is quantized into Landau levels 
(see e.g. Ref.~\cite{Landau:1977})
with energies
\begin{equation}
E(n_L,\,k_{pz})=\frac{k_{pz}^2}{2m_p}+\left(n_L+\frac{1}{2}\right)
\frac{eB}{m_p},\ n_L=0,\ 1,\ 2,\ \cdots,
\end{equation}
where $k_{pz}$ is the proton momentum in the $z$-direction. 
As $eB/m_p=63(B/10^{16}\ \mathrm{G})$ keV, a proton in a gas of temperature
$T\gtrsim 1$ MeV can occupy Landau levels with $n_L\gg 1$ for
$B\sim 10^{16}$ G. From the 
correspondence principle, the proton motion in this case can be considered 
as classical. Thus, we only need to take into account polarization of 
the spin by the magnetic field for both $p$ and $n$.

For the conditions of interest here, $e^-$ and $e^+$ are relativistic. 
Their Landau levels \cite{Johnson:1949} have energies
\begin{equation}
E_e(n,\,p_{ez})=\sqrt{p_{ez}^2+m_e^2+2neB},\ n=0,\ 1,\ 2,\ \cdots,
\label{eq:epz}
\end{equation}
where $p_{ez}$ is the momentum of $e^-$ or $e^+$ in the $z$-direction. 
The result in Eq.~\eqref{eq:epz} takes spin into account. Note
that the Landau levels of $e^-$ and $e^+$ have degeneracy $g_n=1$ 
(corresponding to a single spin state) for the $n=0$ state but $g_n=2$ 
(corresponding to two spin states) for all $n>0$ states.
For a given $E_e$, the maximum value $n_\mathrm{max}$ of $n$ is
\begin{equation}
n_\mathrm{max} = \left[\frac{E_e^2-m_e^2}{2eB}\right]_\mathrm{int}
=\left[8.45\times 10^{-3}\left(\frac{E_e^2-m_e^2}{\mathrm{MeV}^2}\right)
\left(\frac{10^{16}\ \mathrm{G}}{B}\right)\right]_\mathrm{int},
\label{eq:nmax}
\end{equation}
where [\ ]$_\mathrm{int}$ denotes the integer part of the argument.
Thus, $e^-$ and $e^+$ with $E_e\lesssim 10$ MeV can only occupy the
Landau level with $n=0$ in magnetic fields of $B\sim 10^{16}$ G. 
The phase space of $e^-$ and $e^+$ in this case is dramatically 
affected by magnetic fields. In general, the integration over the 
phase space is changed 
from the classical case to the case of Landau levels according to
\begin{equation}
\frac{1}{4\pi^3}\int \ud\Omega\int p_e^2\,\ud p_e\to
\frac{eB}{2\pi^2}\sum_{n=0}^{n_\mathrm{max}}g_n\int \ud p_{ez},
\label{eq:phsp}
\end{equation}
where $\Omega$ is the solid angle in the classical momentum space
and $p_{ez}$ is restricted to positive values. 

\subsection{Heating Rate in Magnetic Fields}
We first calculate the heating rate due to the forward processes in 
Eqs.~\eqref{eq:nun} and \eqref{eq:nup} in magnetic fields.
As we only consider magnetic fields of $B\sim 10^{16}$ G,
the energy scale $\sqrt{eB}=7.69(B/10^{16}\ \mathrm{G})^{1/2}$ MeV is much 
lower than the mass of the $W$ (80 GeV) or $Z$ boson (91 GeV). Thus, the
weak interaction is unaffected by such magnetic fields. However,
such strong magnetic fields can change the cross sections
for the forward processes in Eqs.~\eqref{eq:nun} and \eqref{eq:nup} by
polarizing the spin of $n$ and $p$ in the initial state and by changing 
the phase space of the $e^-$ and $e^+$ in the final state. The new cross 
sections to zeroth order in $m_N^{-1}$ are derived in 
Appendix \ref{sec:cross-sections} using
the Landau wavefunctions of $e^-$ and $e^+$. The results are
\begin{equation}
\sigma_{\nu N}(B)=\sigma_B^{(1)}\left[1+
2\chi\frac{(f\pm g)g}{f^2+3g^2}\cos\Theta_\nu\right]
+\sigma_B^{(2)}\left[\frac{f^2-g^2}{f^2+3g^2}\cos\Theta_\nu
+2\chi\frac{(f\mp g)g}{f^2+3g^2}\right],
\label{eq:xnub}
\end{equation}
where we have factored out two energy-dependent terms
\begin{eqnarray}
\sigma_B^{(1)}&=&\frac{G_{F}^{2}\cos^{2}\theta_{C}}{2\pi}(f^2+3g^2)eB
\sum _{n=0}^{n_\mathrm{max}}\frac{g_nE_e}{\sqrt{E_e^2-m_e^2-2neB}}, \\
\sigma_B^{(2)}&=&\frac{G_{F}^{2}\cos^{2}\theta_{C}}{2\pi}(f^2+3g^2)eB
\frac{E_e}{\sqrt{E_e^2-m_e^2}}.
\end{eqnarray}
In Eq.~\eqref{eq:xnub}, $\Theta_\nu$ is the angle between the neutrino 
momentum and the magnetic field, the upper sign is for $\nu_e$ absoprtion 
on $n$, and the lower sign is for $\bar\nu_e$ absoprtion on $p$. As in 
the case of no magnetic field, $E_e=E_\nu\pm\Delta$.

\begin{figure*}
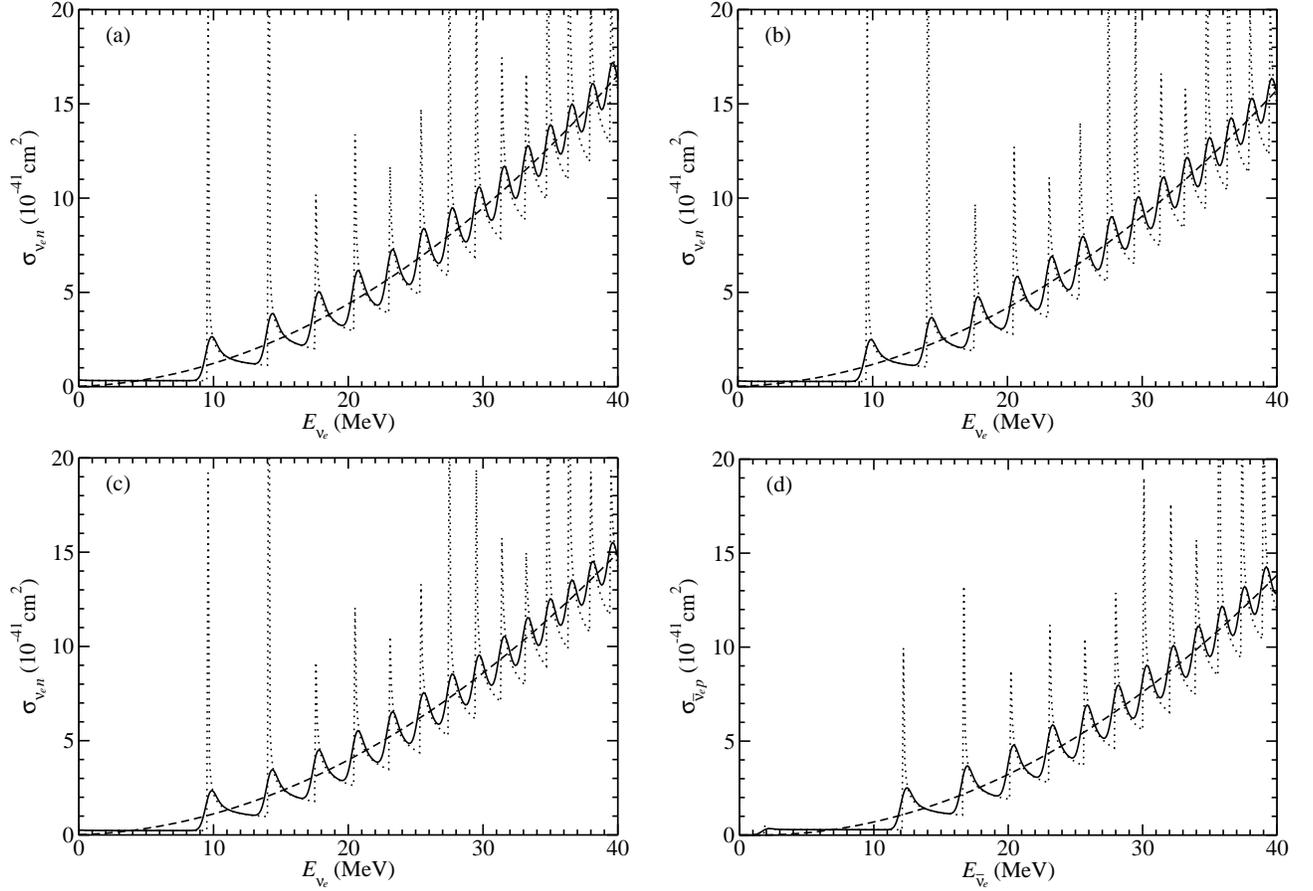

\begin{center}
$\begin{array}{@{}c@{\hspace{0.2in}}c@{}}
\includegraphics*[width=3.25in, keepaspectratio]{fig5a.eps} &
\includegraphics*[width=3.25in, keepaspectratio]{fig5b.eps} \\
\includegraphics*[width=3.25in, keepaspectratio]{fig5c.eps} &
\includegraphics*[width=3.25in, keepaspectratio]{fig5d.eps}
\end{array}$
\end{center}
\caption{\label{fig:5}The cross sections $\sigma_{\nu_en}(B)$ 
in a magnetic field of
$B=10^{16}$ G (dotted curve) and $\sigma_{\nu_en}(B=0)$ (dashed curve)
as functions of $\nu_e$ energy $E_{\nu_e}$ for different angles
$\Theta_{\nu_e}$ between the directions of the magnetic field and the
$\nu_e$ momentum: (a) $\cos\Theta_{\nu_e}=-1$, (b) $\cos\Theta_{\nu_e}=0$,
and (c) $\cos\Theta_{\nu_e}=1$. A neutron polarization of $\chi=-0.05$ is
assumed for all cases. The cross section $\sigma_{\nu_en}(B)$
smoothed with a Gaussian window function
$\exp[-(E-E_{\nu_e})^2/(0.5\ {\rm MeV})^2]$ (solid curve) oscillates
rather symmetrically around $\sigma_{\nu_en}(B=0)$ (dashed curve) at
$E_{\nu_e}\gtrsim 20$ MeV. (d) The cross sections $\sigma_{\bar\nu_ep}(B)$
in a magnetic field of $B=10^{16}$ G (dotted curve) and
$\sigma_{\bar\nu_ep}(B=0)$ (dashed curve) as functions of
$\bar\nu_e$ energy $E_{\bar\nu_e}$ for $\cos\Theta_{\bar\nu_e}=0$.
The dependence on $\cos\Theta_{\bar\nu_e}$ is very weak for the assumed
proton polarization of $\chi=0.05$. The cross section
$\sigma_{\bar\nu_ep}(B)$
smoothed with the same window function (solid curve) exhibits the same
behavior as in the case of $\sigma_{\nu_en}(B)$.}
\end{figure*}

The angular dependence in Eq.~\eqref{eq:xnub} is due to parity violation
of the weak interaction. The dominant angular dependence is associated
with polarization of the spin of the $n$ or $p$ in the initial state.
The single spin state corresponding to the $n=0$ Landau level of the
$e^-$ or $e^+$ in the final state introduces additional (typically
small) angular dependence. For comparison, if there is no magnetic 
field but the spin of the $n$ or $p$ in the initial state is polarized,
the cross sections for the forward processes in 
Eqs.~\eqref{eq:nun} and \eqref{eq:nup} are
\begin{equation}
\sigma_{\nu N}(B=0)=\sigma_{\nu N}^{(0)}
\left[1+2\chi\frac{(f\pm g)g}{f^2+3g^2}\cos\Theta_\nu\right].
\label{eq:xnuchi}
\end{equation}
In Eq.~\eqref{eq:xnuchi}, 
$\sigma_{\nu N}^{(0)}=(G_{F}^{2}\cos^{2}\theta_{C}/\pi)(f^2+3g^2)
p_eE_e$ are the appropriate cross sections for $B=0$ and $\chi=0$ as
given in Eq.~\eqref{eq:sigh}, the upper sign is for $\nu_e$ absoprtion 
on $n$, and the lower sign is for $\bar\nu_e$ absoprtion on $p$.
Note that the angular dependence for
$\bar\nu_e$ absorption on $p$ is much weaker than that for $\nu_e$ 
absorption on $n$ due to the close numerical values of $f$ and $g$.
For illustration, we take $\chi=-0.05$ ($\mu=-1.91\mu_N$ for $n$), 
$B=10^{16}$ G, and calculate 
$\sigma_{\nu_en}(B)$ and $\sigma_{\nu_en}(B=0)$ as functions of 
$E_{\nu_e}$. The results are shown
as the dotted [$\sigma_{\nu_en}(B)$] and dashed [$\sigma_{\nu_en}(B=0)$]
curves in Figs.~\ref{fig:5}a--c for 
$\cos\Theta_{\nu_e}=-1$, 0, and 1, respectively.
For $\chi=0.05$ ($\mu=2.79\mu_N$ for $p$) and $B=10^{16}$ G, the angular 
dependence of $\sigma_{\bar\nu_ep}(B)$ and $\sigma_{\bar\nu_ep}(B=0)$ is 
insignificant. The results for $\cos\Theta_{\bar\nu_e}=0$ are shown
as the dotted [$\sigma_{\bar\nu_ep}(B)$] and dashed 
[$\sigma_{\bar\nu_ep}(B=0)$] curves in Fig.~\ref{fig:5}d.

The cross sections $\sigma_{\nu_en}(B)$ and $\sigma_{\bar\nu_ep}(B)$
shown as the dotted curves in Figs.~\ref{fig:5}a--d 
have spikes superposed on a 
smooth general trend. The
varying heights of these spikes are artifacts of the plotting tool:
all the spikes should have been infinitely high as they correspond to
``resonances'' at $E_e=\sqrt{m_e^2+2neB}$, for which a new Landau level 
opens up. These formal infinities disappear when nucleon motion is taken
into account \cite{dq}. In practice, these formal infinities are 
effectively smoothed out when
integrated over the neutrino energy spectra. To see the behavior of
$\sigma_{\nu N}(B)$ as functions of $E_\nu$ more clearly, we smooth
$\sigma_{\nu N}(B)$ with a Gaussian window function
$\exp[-(E-E_\nu)^2/(0.5\ \mathrm{MeV})^2]$. The results are shown as the 
solid curves in Figs.~\ref{fig:5}a--d. It can be seen that
the smoothed $\sigma_{\nu_en}(B)$ and $\sigma_{\bar\nu_ep}(B)$ oscillate
rather symmetrically around the corresponding results for $B=0$ at 
$E_\nu\gtrsim 20$ MeV. 
This is because for these neutrino energies, the $e^-$ or $e^+$ in the
final state can occupy Landau levels with $n$ up to $n_\mathrm{max}\gtrsim 3$
for $B=10^{16}$ G [see Eq.~\eqref{eq:nmax}]. For $n_\mathrm{max}\gg 1$,
\begin{equation}
eB\sum_{n=0}^{n_\mathrm{max}}\frac{E_{e}}{\sqrt{E_{e}^{2}-m_{e}^{2}-2neB}}
\to\int_0^{\frac{E_e^2-m_e^2}{2}}
\frac{E_{e}}{\sqrt{E_{e}^{2}-m_{e}^{2}-2neB}}\,\ud(neB)
=E_{e}\sqrt{E_{e}^{2}-m_{e}^2},
\end{equation}
and $\sigma_B^{(1)}$ in the dominant term of $\sigma_{\nu N}(B)$
approaches $\sigma_{\nu N}^{(0)}$.

\begin{figure*}
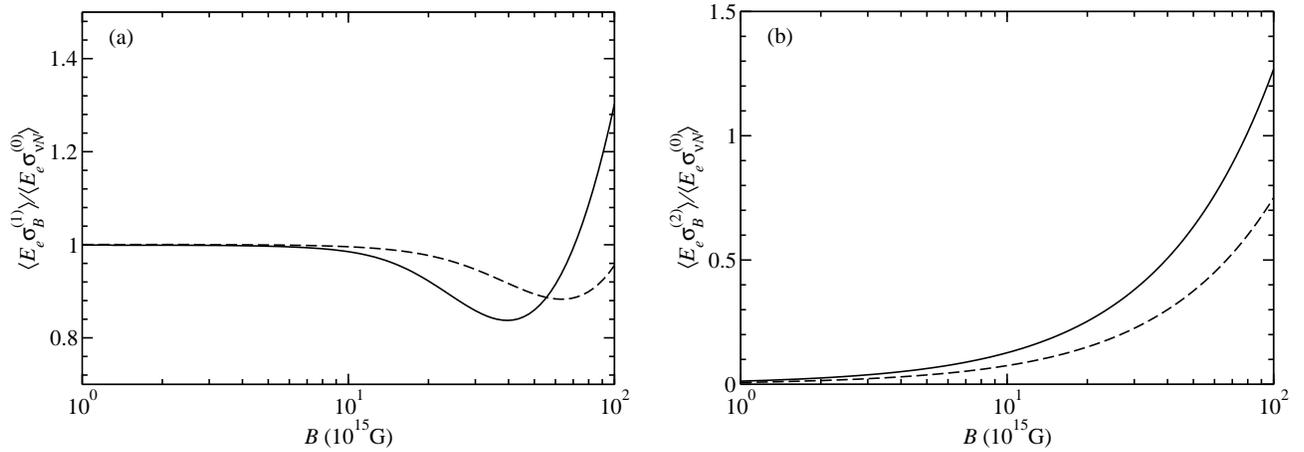

\begin{center}
$\begin{array}{@{}c@{\hspace{.2in}}c@{}}
\includegraphics*[width=3.25in, keepaspectratio]{fig6a.eps} &
\includegraphics*[width=3.25in, keepaspectratio]{fig6b.eps}
\end{array}$
\end{center}
\caption{\label{fig:6}The ratios $\langle E_e\sigma_B^{(1)}\rangle/
\langle E_e\sigma_{\nu N}^{(0)}\rangle$
(a) and $\langle E_e\sigma_B^{(2)}\rangle/\langle E_e\sigma_{\nu N}^{(0)}
\rangle$ (b) as functions of $B$. The solid curve is for $\nu_e$
absorption on $n$ and the dashed curve for $\bar\nu_e$ absorption on $p$.}
\end{figure*}

To calculate the heating rate per nucleon $\dot q_h(B)$
in magnetic fields, we
replace $\sigma_{\nu N}^{(0)}$ with $\sigma_{\nu N}(B)$ in Eq.~\eqref{eq:qh}.
As noted above, the resonances in $\sigma_{\nu N}(B)$ are
smoothed out when integrated over the neutrino energy spectra.
This integration results in
\begin{eqnarray}
\langle E_e\sigma_B^{(1)}\rangle&\equiv&\int E_e\sigma_B^{(1)}
f_\nu(E_\nu)\,\ud E_\nu,\\
\langle E_e\sigma_B^{(2)}\rangle&\equiv&\int E_e\sigma_B^{(2)}
f_\nu(E_\nu)\,\ud E_\nu,
\end{eqnarray}
which can be compared with
\begin{equation}
\langle E_e\sigma_{\nu N}^{(0)}\rangle\equiv\int E_e\sigma_{\nu N}^{(0)}
f_\nu(E_\nu)\,\ud E_\nu.
\end{equation}
The ratios 
$\langle E_e\sigma_B^{(1)}\rangle/\langle E_e\sigma_{\nu N}^{(0)}\rangle$
are shown as functions of $B$ for $\nu_e$ absorption on $n$ (solid curve)
and $\bar\nu_e$ absorption on $p$ (dashed curve) in Fig.~\ref{fig:6}a. The 
corresponding results for 
$\langle E_e\sigma_B^{(2)}\rangle/\langle E_e\sigma_{\nu N}^{(0)}\rangle$
are shown in Fig.~\ref{fig:6}b. It can be seen that for $B\lesssim 10^{16}$ G,
$\langle E_e\sigma_B^{(1)}\rangle/\langle E_e\sigma_{\nu N}^{(0)}\rangle$
stays close to unity. This is because the dominant contributions to the
relevant integrals come from $\nu_e$ with $E_{\nu_e}\sim 20$ MeV or 
$\bar\nu_e$ with $E_{\bar\nu_e}\sim 25$ MeV and these neutrino energies 
correspond to $n_\mathrm{max}\gg 1$. The ratio 
$\langle E_e\sigma_B^{(2)}\rangle/\langle E_e\sigma_{\nu N}^{(0)}\rangle$ 
is negligible for $B\lesssim 10^{16}$ G. In general, the overall term 
involving $\sigma_B^{(2)}$ in $\sigma_{\nu N}(B)$ is much smaller than 
the one involving $\sigma_B^{(1)}$ [see Eq.~\eqref{eq:xnub}].

\begin{figure}
\includegraphics*[width=3.25in, keepaspectratio]{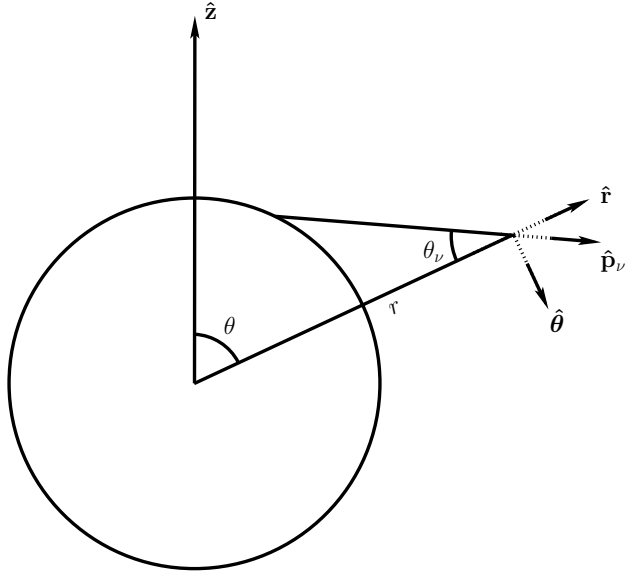}
\caption{\label{fig:7}A sketch of the geometry for integration over 
the neutrino
solid angle to obtain the heating rate per nucleon $\dot q_h(B)$ in
magnetic field. The magnetic field is uniform and in the $z$-direction.
The cross sections used to calculate $\dot q_h(B)$ depend on the angle
$\Theta_\nu$ between the directions of the magnetic field
($\mathbf{\hat z}$) and the neutrino momentum ($\mathbf{\hat p}_\nu$).
The integration over the neutrino solid angle can be performed by
expressing $\cos\Theta_\nu=\mathbf{\hat z\cdot\hat p}_\nu$ in terms of
$\theta$, $\theta_\nu$, and $\phi_\nu$ (the azimuthal neutrino angle
$\phi_\nu$ corresponding to $\theta_\nu$ is not shown).}
\end{figure}

It can be seen from Figs.~\ref{fig:6}a and 6b that 
substantial changes to the magnitude of the heating rate only
occur for $B\sim 10^{17}$ G. However, a qualitatively new and 
quantitatively significant effect already occurs for $B\sim 10^{16}$ G.
Due to polarization of the spin of $n$ or $p$ in the initial state of 
the heating reactions, the heating rate at position $\bm{r}$ depends on
the angle $\theta$ between $\bm{r}$ and the magnetic field (in the
$z$-direction). This angular dependence enters through integration
over the neutrino solid angle (see Fig.~\ref{fig:7})
\begin{equation}
\int\cos\Theta_\nu \,\ud\Omega_\nu=\pi(R_\nu/r)^2\cos\theta.
\end{equation}

\subsection{Cooling Rate in Magnetic Fields}
Next we calculate the cooling rate due to the reverse processes in 
Eqs.~\eqref{eq:nun} and \eqref{eq:nup} in magnetic fields.
The differential volume reaction rates $\ud\Gamma_{eN}(B)/\ud\!\cos\Theta_\nu$
for these processes are derived in Appendix \ref{sec:cross-sections}. 
The results are
\begin{equation}
\frac{\ud\Gamma_{eN}(B)}{\ud\!\cos\Theta_\nu}=\frac{\Gamma_{eN}^{(0)}}{2}
\left[1+2\chi\frac{(f\pm g)g}{f^2+3g^2}\cos\Theta_\nu
+\frac{f^2-g^2}{f^2+3g^2}\cos\Theta_\nu
+2\chi\frac{(f\mp g)g}{f^2+3g^2}\right]
\label{eq:geb0}
\end{equation}
if the $e^-$ or $e^+$ in the initial state is in the $n=0$ Landau 
level, and
\begin{equation}
\frac{\ud\Gamma_{eN}(B)}{\ud\!\cos\Theta_\nu}=\frac{\Gamma_{eN}^{(0)}}{2}
\left[1+2\chi\frac{(f\pm g)g}{f^2+3g^2}\cos\Theta_\nu\right]
\label{eq:gebn}
\end{equation}
if the $e^-$ or $e^+$ in the initial state is in the $n>0$ Landau level.
In Eqs.~\eqref{eq:geb0} and \eqref{eq:gebn},
$\Gamma_{eN}^{(0)}=[G_F^2\cos^2\theta_C/(2\pi)](f^2+3g^2)E_\nu^2$
with $E_\nu=E_e\pm\Delta$ is the volume reaction rate in the absence of 
magnetic field as given in Eq.~\eqref{eq:sigc}, the upper sign is for
$e^+$ capture on $n$, and the lower sign is for $e^-$ capture on $p$.
The dependence of $\ud\Gamma_{eN}(B)/\ud\!\cos\Theta_\nu$ on the direction
of the neutrino emitted in the final state again manifests parity
violation of the weak interaction. As we are not interested in the 
neutrinos emitted by the cooling processes, we integrate over 
$\cos\Theta_\nu$ to obtain the volume reaction rate
\begin{equation}
\Gamma_{eN}(B)=\left\{\begin{array}{ll}
\Gamma_{eN}^{(0)}[1+2\chi(f\mp g)g/(f^2+3g^2)],&n=0,\\
\Gamma_{eN}^{(0)},&n>0.\end{array}\right.
\end{equation}
Clearly, the volume reaction rates of the cooling processes are not much 
affected by the magnetic field for $|\chi|\ll 1$.

However, magnetic fields also affect the cooling rate through the 
equations of state for $e^-$ and $e^+$ (see Appendix \ref{sec:eos}).
For a given set of $Y_e$, $S$, and $T$, the density $\rho$ and 
the electron degeneracy parameter $\eta_e$ 
in the presence of magnetic field differ
from those in the case of no magnetic field. 
Taking $Y_e=0.5$ and $T(r)$ in Eq.~\eqref{eq:tr}, we show $\rho(r)$
for $B=10^{16}$ G as the dot-dashed ($S=10$) and dot-dot-dashed
($S=20$) curves along with the corresponding results for $B=0$
[solid ($S=10$) and dashed ($S=20$) curves] in Fig.~\ref{fig:4}a. 
The comparison
for $\eta_e(r)$ is shown in Fig.~\ref{fig:4}b.
It can be seen that for the same $Y_e$ and $T$,
magnetic fields of $B\sim 10^{16}$ G change (mostly decrease) $\rho$ 
slightly but decrease $\eta_e$ greatly for $S=10$.
The same magnetic fields significantly increase $\rho$ but decrease 
$\eta_e$ for $S=20$. Note that $\eta_e$ is already small for $B=0$
and $S=20$.

\begin{figure*}
\begin{center}
$\begin{array}{@{}c@{\hspace{.2in}}c@{}}
\includegraphics*[width=3.25in, keepaspectratio]{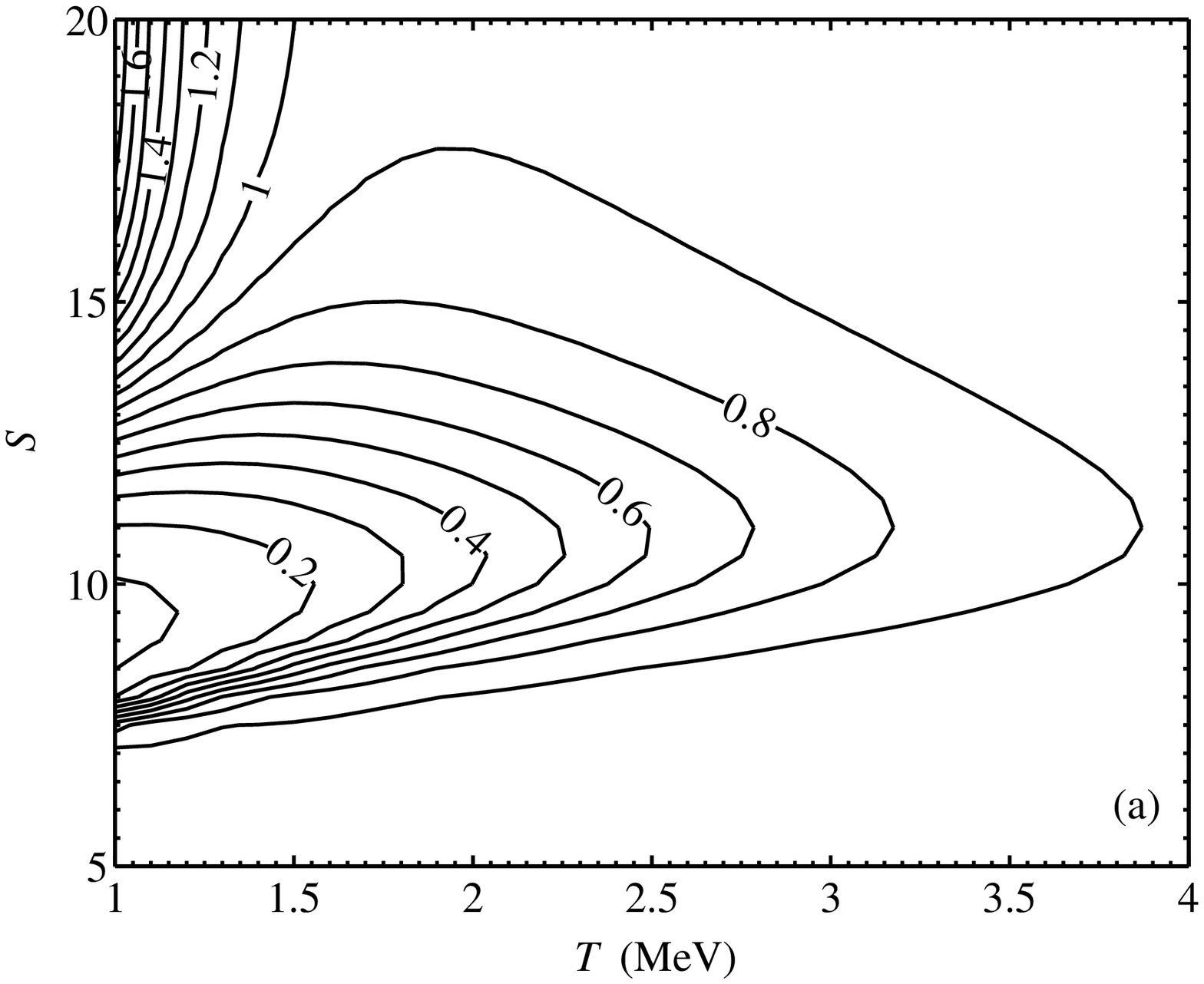} &
\includegraphics*[width=3.25in, keepaspectratio]{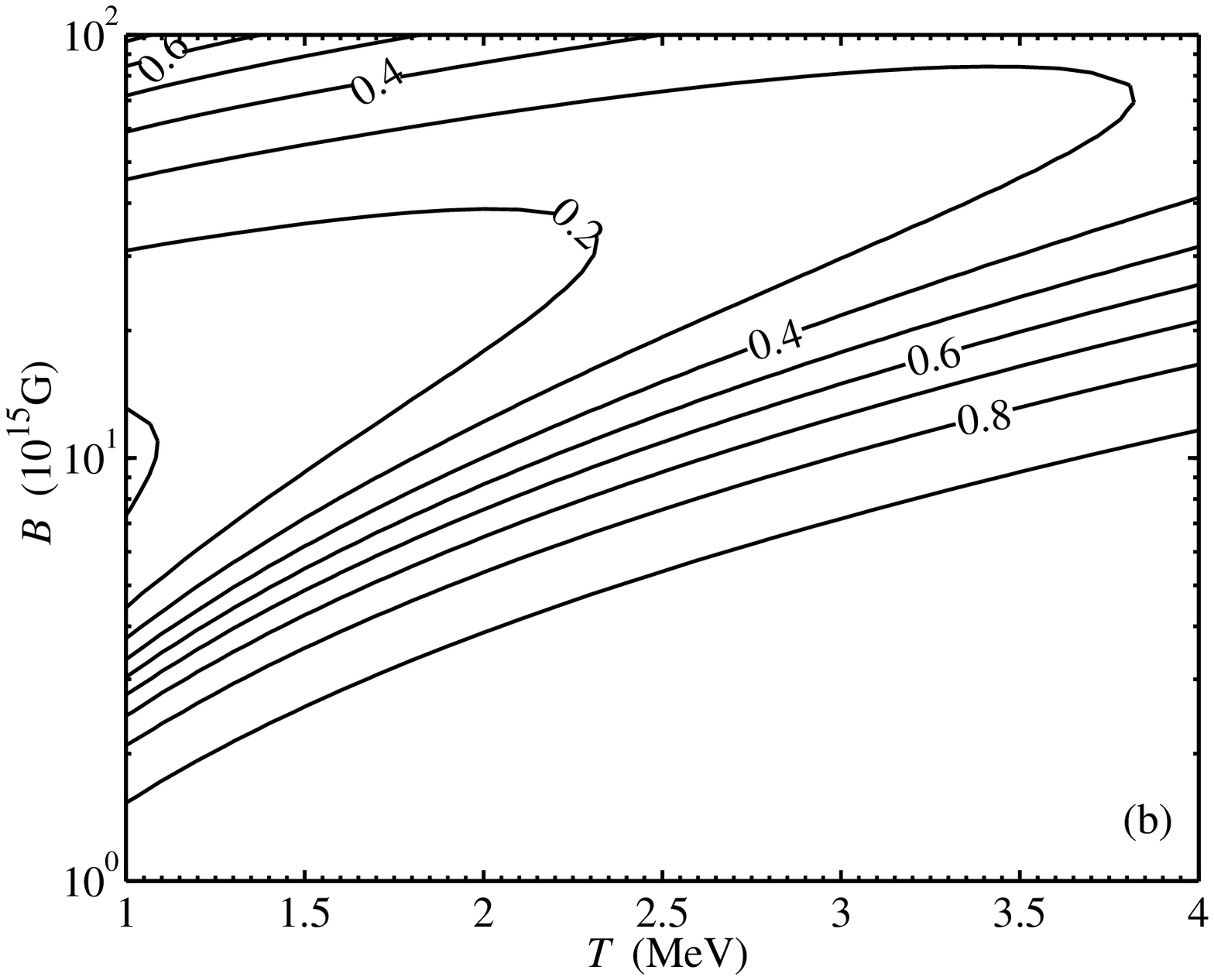}
\end{array}$
\end{center}
\caption{\label{fig:8}Contours of constant $\dot q_c(B)/\dot q_c^{(0)}$ 
in increment
of 0.1 as functions of $S$ and $T$ for $B=10^{16}$ G (a) and as functions
of $B$ and $T$ for $S=10$ (b). Magnetic fields of $B\sim 10^{16}$ G
greatly reduce the cooling rate per nucleon for entropies of $S\sim 10$,
especially at temperatures of $T\lesssim 2$ MeV.}
\end{figure*}

The cooling rate per nucleon in magnetic fields is
\begin{eqnarray}
\dot q_c(B)&=&Y_n\frac{eB}{2\pi^2}\sum_{n=0}^\infty g_n\int_0^\infty
\frac{E_{e^+}\Gamma_{e^+n}(B)}{\exp[(E_{e^+}/T)+\eta_e]+1}\,\ud p_{e^+z}
\nonumber\\
 & & +Y_p\frac{eB}{2\pi^2}\sum_{n=0}^\infty g_n\int_{p_{e^-z,n}}^\infty
\frac{E_{e^-}\Gamma_{e^-p}(B)}{\exp[(E_{e^-}/T)-\eta_e]+1}\,\ud p_{e^-z}.
\label{eq:qcb}
\end{eqnarray}
In Eq.~\eqref{eq:qcb}, the threshold momentum $p_{e^-z,n}$ for $e^-$ capture
on $p$ corresponds to $p_{e^-z,n}^2=\Delta^2-m_e^2-2neB$ or 0, whichever
is larger. Taking $Y_e=0.5$, we plot contours of constant
$\dot q_c(B)/\dot q_c^{(0)}$ as functions of $S$ and $T$ for 
$B=10^{16}$ G in Fig.~\ref{fig:8}a and as functions of $B$ and $T$ for $S=10$
in Fig.~\ref{fig:8}b. It can be seen that magnetic fields of
$B\sim 10^{16}$ G can decrease the cooling rate significantly
for $S\sim 10$. This is mostly due to the reduction of $\eta_e$
through the effects of magnetic fields
on the equations of state for $e^-$ and $e^+$ (see Fig.~\ref{fig:4}b).

\section{\label{sec:implications}Implications for Supernova Dynamics}
Now we consider the effects of strong magnetic fields on supernova 
dynamics using the heating and cooling rates discussed in 
Sec.~\ref{sec:rates-in-fields}.
As mentioned earlier, observations suggest that magnetic fields of at
least $\sim 10^{15}$ G can be generated during the formation of 
protoneutron stars. However, little is known about the actual strength
and topology of protoneutron star magnetic fields. To illustrate the
potential effects of such fields on supernova dynamics, we consider
simple cases of uniform and dipole fields of $\sim 10^{16}$ G.

\subsection{Uniform Field}
We first consider the case of a uniform magnetic field in the 
$z$-direction. The heating and cooling rates in such a field have been
discussed in Sec.~\ref{sec:rates-in-fields}. 
For illustration, we take $B=10^{16}$ G.
The corresponding heating rate is
\begin{eqnarray}
\dot q_h(B=10^{16}\ \mathrm{G})&=&\{3.77\times 10^3Y_n
[1-1.46\times 10^{-2}\chi_n+(0.988\chi_n-1.31\times 10^{-2})
\Phi(r)\cos\theta]\nonumber\\
& & +5.29\times 10^3Y_p
[1+7.45\times 10^{-2}\chi_p-(0.114\chi_p+7.69\times 10^{-3})
\Phi(r)\cos\theta]\}\nonumber\\
& & \times\left[1-\sqrt{1-(R_\nu/r)^2}\right]\ 
\mathrm{MeV\ s}^{-1}\ \mathrm{nucleon}^{-1},
\label{eq:qhb}
\end{eqnarray} 
where $\chi_n$ and $\chi_p$ are the net polarization of $n$ and $p$,
respectively, as given in Eq.~\eqref{eq:chi}, and
\begin{equation}
\Phi(r)=\frac{(R_\nu/r)^2}{2\left[1-\sqrt{1-(R_\nu/r)^2}\right]}.
\end{equation}
Note that $\Phi(r)$ increases from 1/2 to 1 as $r$ increases from $R_\nu$
to $r\gg R_\nu$. Note also that $\chi_n$ and $\chi_p$ are functions of $r$
through their dependence on $T(r)$ 
[see Eqs.~\eqref{eq:tr} and \eqref{eq:chi}]. 
For $T\sim 2$ MeV, $\chi_n\sim -0.03$ and $\chi_p\sim 0.04$. The magnitudes 
of $\chi_n$ and $\chi_p$ increase for lower $T$. Thus, in the region of
$T\lesssim 2$ MeV [$r\gtrsim 100$ km, see Eq.~\eqref{eq:tr}], 
the heating rate in Eq.~\eqref{eq:qhb} 
varies by at least several percent over $-1\leq\cos\theta\leq 1$
for a given $r$. This variation can
induce or amplify anisotropy in the bulk motion of the material below the 
stalled shock, eventually producing an asymmetric explosion. 
The protoneutron star would then receive a ``kick'' during the explosion.
Assuming that $\sim 1\,M_\odot$
of material with a kinetic energy of $\sim 10^{51}$ erg is below the
shock when the explosion starts, a $\sim 2\%$ asymmetry in the bulk 
motion of this material would result in a kick velocity of
$\sim 0.02(10^{51}\ \mathrm{erg}/1\,M_\odot)^{1/2}\sim 140$ km s$^{-1}$ for
the protoneutron star. This could explain the observed velocities for
a large fraction of pulsars \cite{Cordes:1998}.

\begin{figure}
\includegraphics*[width=3.25in, keepaspectratio]{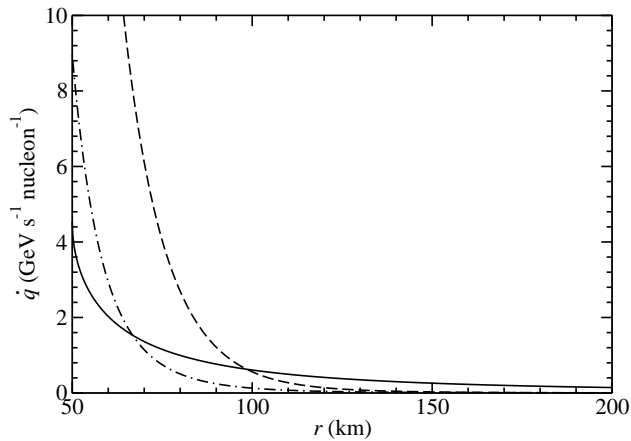}
\caption{\label{fig:9}The angle-averaged heating rate per 
nucleon (solid curve) and
the cooling rate per nucleon [dashed ($S=10$) and dot-dashed ($S=20$)
curves] in a uniform magnetic field of $B=10^{16}$ G as functions of
radius $r$. Compared with Fig.~\ref{fig:3}, 
the gain radius decreases significantly
from 137 km for $B=0$ to 99 km for $B=10^{16}$ G in the case of $S=10$
but essentially remains at 68 km in the case of $S=20$.}
\end{figure}

\begin{figure}
\includegraphics*[width=3.25in, keepaspectratio]{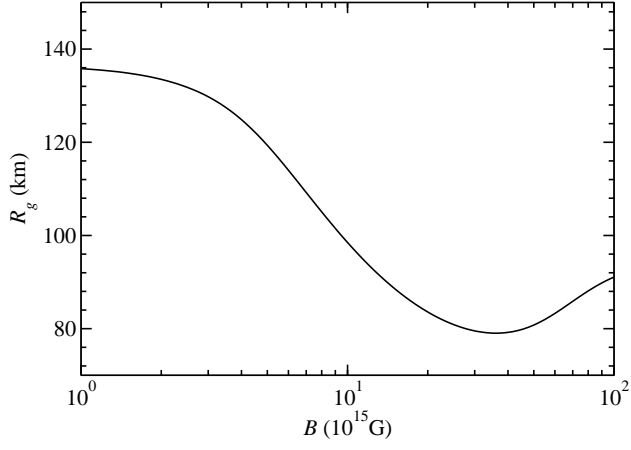}
\caption{\label{fig:10}The gain radius $R_g$ as a function of magnetic field
strength $B$ for a uniform magnetic field ($Y_e=0.5$ and $S=10$). 
The gain radius for $B\sim 10^{15}$ G approaches that for $B=0$.}
\end{figure}

While the angular dependence of the heating rate has some interesting
dynamic effects as discussed above, it only introduces minor perturbation
on the position of the gain radius. To very good approximation, 
one may use the angular
average of the heating rate [obtained effectively by dropping the terms
involving $\cos\theta$ in Eq.~\eqref{eq:qhb}]
in determining the gain radius.
The angle-averaged heating rate for $Y_e=0.5$
is shown as a function of $r$ (solid curve) in Fig.~\ref{fig:9}. 
In the same figure, 
we also show the cooling rate $\dot q_c(B=10^{16}\ \mathrm{G})$ as a function 
of $r$ using $T(r)$ in Eq.~\eqref{eq:tr} and $Y_e=0.5$ for $S=10$ 
(dashed curve) and $S=20$ (dot-dashed curve), respectively. By comparing
Figs.~\ref{fig:3} and \ref{fig:9}, 
it can be seen that the gain radius decreases significantly
from 137 km for $B=0$ to 99 km for $B=10^{16}$ G in the case of
$S=10$ but essentially remains at 68 km in the case of $S=20$. This is 
because the magnetic field greatly reduces $\eta_e$ (see Fig.~\ref{fig:4}b), 
and hence, the cooling rate (see Fig.~\ref{fig:8}a) 
for $S=10$. But for $S=20$, 
$\eta_e$ is already small for $B=0$ (see Fig.~\ref{fig:4}b) 
and reduction of $\eta_e$
by the magnetic field does not change the cooling rate significantly 
(see Fig.~\ref{fig:8}a). Numerical models \cite{Rampp:2000,Liebendorfer:2001} 
show that the material
below the stalled shock initially has $S\sim 10$.
Taking $T(r)$ in Eq.~\eqref{eq:tr}, $Y_e=0.5$, and 
$S=10$, we calculate the gain radius as a function of $B$ and show the
results in Fig.~\ref{fig:10}. It can be seen that magnetic fields of 
$B\sim 10^{16}$ G or larger significantly decrease the gain radius, 
thereby enhancing the net heating below the stalled shock. Consequently,
the shock may be revived more efficiently (i.e., within a shorter 
time) to make an explosion.

\subsection{Dipole Field}
As a second example, we consider the magnetic field of a dipole in
the $z$-direction:
\begin{equation}
\mathbf{B}=B_0\left(\frac{R_\nu}{r}\right)^3
(2\cos\theta\,\mathbf{\hat r}+\sin\theta\,\mathbf{\hat{\bm{\theta}}}).
\end{equation}
The heating and cooling rates in a uniform field discussed in 
Sec.~\ref{sec:rates-in-fields}
can be adapted to the case of a dipole field in a straightforward manner.
The strength of the magnetic field to be used in the expressions for
$\chi$ [Eq.~\eqref{eq:chi}] and $E_e(n,p_{ez})$ [Eq.~\eqref{eq:epz}] is now
\begin{equation}
B(r,\theta)=B_0\left(\frac{R_\nu}{r}\right)^3\sqrt{1+3\cos^2\theta}.
\label{eq:bdipole}
\end{equation}
In addition, the integration over the neutrino solid angle 
(see Fig.~\ref{fig:7})
to obtain the heating rate is changed to
\begin{equation}
\int\cos\Theta_\nu \ud\Omega_\nu=2\pi(R_\nu/r)^2
\frac{\cos\theta}{\sqrt{1+3\cos^2\theta}}.
\end{equation}
\begin{figure}
\includegraphics*[width=3.25in, keepaspectratio]{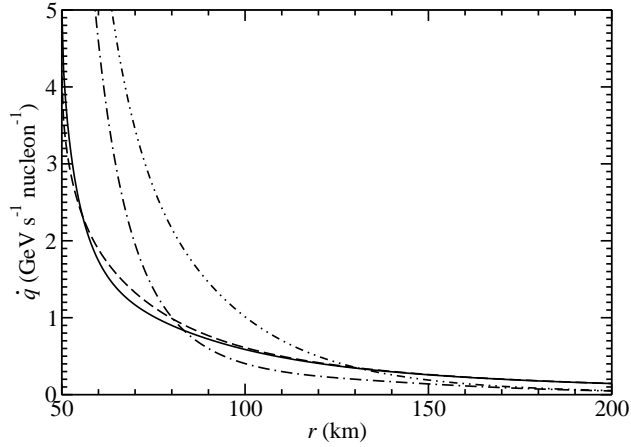}
\caption{\label{fig:11}The heating rate per nucleon [dashed 
($\cos\theta=0$) and
solid ($\cos\theta=1$) curves] and the cooling rate per nucleon
[dot-dot-dashed ($\cos\theta=0$) and dot-dashed ($\cos\theta=1$) curves]
in a dipole magnetic field as functions of radius $r$ 
($Y_e=0.5$ and $S=10$). The dipole field
has strength $B(r,\theta)=B_0(R_\nu/r)^3\sqrt{1+3\cos^2\theta}$ with
$B_0=5\times 10^{16}$ G.}
\end{figure}

\begin{figure}
\includegraphics*[width=3.25in, keepaspectratio]{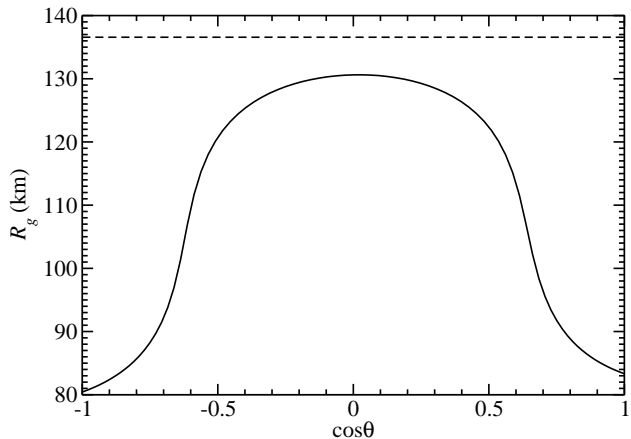}
\caption{\label{fig:12}The gain radius $R_g$ as a function of 
$\cos\theta$
(solid curve) for the dipole magnetic field in Fig.~\ref{fig:11}. Compared
with $R_g=137$ km for $B=0$ (dashed curve), $R_g$ is substantially
reduced to $\sim 80$ km close to the north and south poles of the
magnetic field ($|\cos\theta|\sim 1$).}
\end{figure}

For illustration, we take $B_0=5\times 10^{16}$ G and show the heating 
rate as a function of $r$ for $\cos\theta=0$ (dashed curve) and 
$\cos\theta=1$ (solid curve), respectively,
in Fig.~\ref{fig:11}. In the same figure, we also show the corresponding
cooling rate using $T(r)$ in Eq.~\eqref{eq:tr}, $Y_e=0.5$, and $S=10$
[dot-dot-dashed ($\cos\theta=0$) and dot-dashed ($\cos\theta=1$) curves].
It can be seen that the gain radius differs significantly for
$\cos\theta=0$ and 1. For a close examination, we show the gain radius
$R_g$ as a function of $\cos\theta$ (solid curve) in Fig.~\ref{fig:12}. 
Compared
with $R_g=137$ km for $B=0$ (dashed curve), $R_g$ is substantially
reduced to $\sim 80$ km close to the north and south poles of the
magnetic field ($|\cos\theta|\sim 1$). This reduction in $R_g$ is somewhat
larger than that for the uniform magnetic field discussed above.
This is because the strength of the dipole field at $r\sim 100$ km
in the polar directions is somewhat larger than the strength
of $B=10^{16}$ G taken for the uniform field. By comparison, $R_g$ only
decreases slightly from 137 km for $B=0$ to 131 km near the equator of
the dipole field ($|\cos\theta|\ll 1$). This is because at a given $r$,
the strength of a dipole field for $|\cos\theta|\ll 1$ is weaker than 
that for $|\cos\theta|\sim 1$ by a factor of $\sim 2$
[see Eq.~\eqref{eq:bdipole}].

The gain radius for $|\cos\theta|\sim 1$ is significantly smaller than
that for $|\cos\theta|\ll 1$ in the case of a dipole field. 
This indicates that the explosion is very likely
to occur first in the polar directions of the magnetic field. Note that
the ejecta in the north and south poles tend to kick the protoneutron 
star in opposite directions. However, there would be a net kick as the
heating rate differs for $\cos\theta=-1$ and 1. This can be seen from
Fig.~\ref{fig:12}, which shows a small difference in the gain radius for 
$\cos\theta=-1$ ($R_g=80$ km) and $\cos\theta=1$ ($R_g=83$ km).
Thus, we expect that the possible kick received by the protoneutron star
is similar for both the uniform and the dipole fields discussed here.

\section{\label{sec:conclusions}Conclusions}
We have calculated the rates of heating and cooling due to the neutrino
processes in Eqs.~\eqref{eq:nun} and \eqref{eq:nup} in strong
magnetic fields. We find that for $B\sim 10^{16}$ G, the main effect of 
magnetic fields is to change the
equations of state through the phase space of $e^-$ and $e^+$, which
differs from the classical case due to quantization of the motion of 
$e^-$ and $e^+$ perpendicular to the magnetic field. As a result,
the cooling rate can be greatly reduced by magnetic fields of 
$\sim 10^{16}$ G for typical conditions ($S\sim 10$) below the stalled 
shock and a nonuniform protoneutron star magnetic field (e.g., a
dipole field) can introduce a large angular dependence of the
cooling rate. In addition, strong magnetic fields always lead to an
angle-dependent heating rate by polarizing the spin of $n$ and $p$. 
The decrease in the cooling rate for magnetic fields of 
$\sim 10^{16}$ G decreases the gain radius and increases
the net heating below the stalled shock. We conclude that if magnetic 
fields of $B\sim 10^{16}$ G exist within $\sim 100$ km 
of the protoneutron star, the shock can be revived more efficiently 
(i.e., within a shorter time) to make an explosion.
In addition, the anisotropy in the heating rate induced by strong 
magnetic fields and that in the cooling rate induced by strong nonuniform
(e.g., dipole-like) magnetic fields may lead to significant asymmetry 
in the bulk motion of the material below the stalled shock, eventually 
producing an asymmetric supernova explosion. We speculate that this may 
be one of the mechanisms for producing 
the observed pulsar kick velocities.
Obviously, a full treatment of magnetic fields during the formation of a
protoneutron star and during
the supernova process in general greatly increases
the complexity of an already difficult problem. Nevertheless, we hope
that the interesting effects of strong magnetic fields discussed here
would help to motivate the eventual inclusion of magnetic fields
in supernova models.

\begin{acknowledgments}
We would like to thank Arkady Vainshtein for helpful discussions. 
This work was supported in part by DOE grants DE-FG02-87ER40328
and DE-FG02-00ER4114.
\end{acknowledgments}

\appendix
\section{\label{sec:eos}Equations of State}
The relevant equations of state concern the net electron number density
$(n_{e^-}-n_{e^+})$ [Eq.~\eqref{eq:eosye}] and the total entropy per nucleon 
$S$ [Eq.~\eqref{eq:eoss}]. The photon contribution $S_{\gamma}$ to $S$ is: 
\begin{equation}
S_{\gamma}=\frac{4\pi^{2}}{45}\left(\frac{m_{N}}{\rho}\right)T^{3}
=1.9\left(\frac{T}{\textrm{MeV}}\right)^{3}
\left(\frac{10^{8}\,\textrm{g}\,\textrm{cm}^{-3}}{\rho}\right).
\end{equation}
For the conditions in the region of interest, nucleons are nondegenerate
and nonrelativistic. So their contribution $S_N$ to $S$ is:
\begin{eqnarray}
S_{N} & = & \frac{5}{2}+Y_{n}\ln\left[\frac{2m_{N}}{\rho Y_{n}}
\left(\frac{m_{N}T}{2\pi }\right)^{3/2}\right]
+Y_{p}\ln\left[\frac{2m_{N}}{\rho Y_{p}}
\left(\frac{m_{N}T}{2\pi }\right)^{3/2}\right]\nonumber\\
 & = & 11.5+\ln \left[\left(\frac{T}{\textrm{MeV}}\right)^{3/2}
\left(\frac{10^{8}\,\textrm{g}\,\textrm{cm}^{-3}}{\rho}\right)\right]
-Y_{e}\ln Y_{e}-(1-Y_{e})\ln(1-Y_{e}),
\end{eqnarray}
where we have used $Y_n=1-Y_e$ and $Y_p=Y_e$.
The above expressions for $S_{\gamma}$ and $S_N$ are valid for both the 
case of no magnetic field and the case of magnetic fields of 
$B\sim 10^{16}$ G considered here.

In the absence of magnetic field, the general expressions for the number
densities $n_{e^{\pm }}$, energy densities $\varepsilon_{e^{\pm }}$, and 
pressure $P_{e^{\pm }}$ of $e^+$ and $e^-$ are:
\begin{eqnarray}
n_{e^{\pm }} & = & \frac{1}{\pi^2}\int_0^\infty
\frac{p_{e^{\pm }}^2}
{\exp [(E_{e^{\pm }}/T)\pm \eta _{e}]+1}\,\ud p_{e^{\pm }}\nonumber \\
 & = & \frac{T^{3}}{\pi ^{2}}\int _{0}^{\infty }
\frac{x^2}{\exp \left[\sqrt{x^{2}+(m_{e}/T)^{2}}\pm \eta _{e}\right]+1}
\,\ud x,
\label{eq:num}
\end{eqnarray}
\begin{eqnarray}
\varepsilon_{e^{\pm }}& = & \frac{1}{\pi^2}\int_0^\infty
\frac{E_{e^{\pm }}p_{e^{\pm }}^2}
{\exp [(E_{e^{\pm }}/T)\pm \eta _{e}]+1}\,\ud p_{e^{\pm }}\nonumber \\
 & = & \frac{T^{4}}{\pi ^{2}}\int _{0}^{\infty }
\frac{x^{2}\sqrt{x^{2}+(m_{e}/T)^{2}}}
{\exp \left[\sqrt{x^{2}+(m_{e}/T)^{2}}\pm \eta _{e}\right]+1}\,\ud x,
\label{eq:eden}
\end{eqnarray}
and
\begin{eqnarray}
P_{e^{\pm }} & = & \frac{1}{3\pi ^{2}}\int_0^\infty
\frac{p_{e^{\pm }}^4/E_{e^{\pm }}}
{\exp [(E_{e^{\pm }}/T)\pm \eta _{e}]+1}\,\ud p_{e^{\pm }}\nonumber \\
 & = & \frac{T^{4}}{3\pi ^{2}}\int _{0}^{\infty }
\frac{x^{4}/\sqrt{x^{2}+(m_{e}/T)^{2}}}
{\exp \left[\sqrt{x^{2}+(m_{e}/T)^{2}}\pm \eta _{e}\right]+1}\,\ud x.
\label{eq:pre}
\end{eqnarray}
In Eqs.~\eqref{eq:num}--\eqref{eq:pre}, the upper sign is for $e^+$ and 
the lower sign for $e^-$.

In magnetic fields, the energy levels and the phase space
of $e^+$ and $e^-$ are changed [see Eqs.~\eqref{eq:epz} and \eqref{eq:phsp}].
The corresponding expressions for $n_{e^\pm}$, $\varepsilon_{e^\pm}$, 
and $P_{e^\pm}$ are:
\begin{eqnarray}
n_{e^{\pm }} & = & \frac{eB}{2\pi ^2}\sum _{n=0}^{\infty }g_{n}
\int _{0}^{\infty }\frac{1}{\exp [(E_{e^{\pm }}/T)\pm \eta _{e}]+1}
\,\ud p_{e^\pm z}\nonumber \\
& = & \frac{eBT}{2\pi ^{2}}\sum _{n=0}^{\infty }g_{n}\int _{0}^{\infty }
\frac{1}{\exp 
\left\{\sqrt{x^{2}+[(m_{e}^2+2neB)/T^{2}]}\pm \eta _{e}\right\}+1}\,\ud x,
\label{eq:numb}
\end{eqnarray}
\begin{eqnarray}
\varepsilon_{e^{\pm }}& = & \frac{eB}{2\pi ^2}\sum _{n=0}^{\infty }g_{n}
\int _{0}^{\infty }\frac{E_{e^{\pm }}}
{\exp [(E_{e^{\pm }}/T)\pm \eta _{e}]+1}\,\ud p_{e^\pm z}\nonumber \\
 & = & \frac{eBT^2}{2\pi ^{2}}\sum _{n=0}^{\infty }g_{n}
\int_{0}^{\infty} \frac{\sqrt{x^{2}+[(m_{e}^2+2neB)/T^{2}]}}{\exp 
\left\{\sqrt{x^{2}+[(m_{e}^2+2neB)/T^{2}]}\pm \eta _{e}\right\}+1}\,\ud x,
\label{eq:edenb}
\end{eqnarray}
and
\begin{eqnarray}
P_{e^{\pm }} & = & \frac{eB}{2\pi ^2}\sum _{n=0}^{\infty }g_{n}
\int _{0}^{\infty }\frac{p_{e^\pm z}^{2}/E_{e^{\pm }}}
{\exp [(E_{e^{\pm }}/T)\pm \eta _{e}]+1}\,\ud p_{e^\pm z}\nonumber \\
 & = & \frac{eBT^2}{2\pi ^{2}}\sum _{n=0}^{\infty }g_{n}
\int _{0}^{\infty }
\frac{x^2/\sqrt{x^{2}+[(m_{e}^2+2neB)/T^{2}]}}
{\exp \left\{\sqrt{x^{2}+[(m_{e}^2+2neB)/T^{2}]}\pm \eta _{e}\right\}+1}
\,\ud x.
\label{eq:preb}
\end{eqnarray}
In Eqs.~\eqref{eq:numb}--\eqref{eq:preb}, the upper sign is for $e^+$ and 
the lower sign for $e^-$.

The contributions $S_{e^\pm}$ from $e^+$ and $e^-$ to $S$ can be obtained
in terms of $n_{e^\pm}$, $\varepsilon_{e^\pm}$, and $P_{e^\pm}$ as
\begin{equation}
S_{e^{\pm }} = \left(\frac{m_{N}}{\rho }\right)
\left[\frac{\varepsilon_{e^{\pm }}+P_{e^{\pm }}}{T}
\pm \eta _{e}n_{e^{\pm }}\right].
\end{equation}
More specifically,
\begin{equation}
S_{e^-}+S_{e^+} = 
\frac{\varepsilon_{e^-}+\varepsilon_{e^+}+P_{e^-}+P_{e^+}}{T(\rho/m_{N})}
-\eta _{e}Y_e,
\end{equation}
where we have used $n_{e^-}-n_{e^+}=Y_e\rho/m_{N}$.

\section{\label{sec:cross-sections}
Neutrino processes in strong magnetic fields}

A number of studies on neutrino processes in strong magnetic fields 
exist in the literature. The forward and reverse processes in 
Eq.~\eqref{eq:nun} have been studied in Refs.~\cite{Chandra:2002} and
\cite{Leinson:1998} assuming that $e^-$ and $p$ are in the ground
$(n=0)$ Landau levels. All the four processes in Eqs.~\eqref{eq:nun}
and \eqref{eq:nup} have been studied in Ref.~\cite{Gvozdev:1999}
assuming that $e^-$ and $e^+$ only occupy the ground Landau levels.
The forward processes in Eqs.~\eqref{eq:nun} and \eqref{eq:nup}
have been studied in Refs.~\cite{Roulet:1998sw} and \cite{Lai:1998sz}
assuming that the magnetic field only affects the phase space of 
$e^-$ and $e^+$. Parity violation in the forward process in Eq.~\eqref{eq:nun}
has been studied in Ref.~\cite{Arras:1998mv}. The cross section of the
forward process in Eq.~\eqref{eq:nun} has been calculated in 
Ref.~\cite{Bhattacharya:2002qf} using an approach similar to ours.
In this appendix, we treat the forward and reverse processes in 
Eqs.~\eqref{eq:nun} and \eqref{eq:nup} in magnetic fields of 
$B \sim 10^{16}\,\textrm{G}$. Such fields are strong enough to change the 
motion of $e^-$ and $e^+$ but do not affect the description of weak interation
(see Sec.~\ref{sec:rates-in-fields}).

Ignoring higher order corrections, we take
the effective four-fermion Lagrangian of weak interaction to be
\begin{equation}
\mathcal{L}_{\mathrm{int}} = \frac{G_{F}}{\sqrt{2}}
\cos \theta_{C} N_{\alpha}^{\dagger} L^{\alpha} +\mathrm{h.c.},
\label{eq:lagrangian}
\end{equation}
where $\mathrm{h.c.}$ means the Hermitian conjugation of the first term.
In Eq.~\eqref{eq:lagrangian},
the leptonic charged current $L^\alpha$ has the classical $V-A$ form
\begin{equation}
L^{\alpha} = \bar{\psi}^{e} \gamma^{\alpha} (1 - \gamma_{5}) \psi^{\nu},
\end{equation}
and the nucleonic current $N^\alpha$ is
\begin{equation}
N^{\alpha} = \bar{\psi}^{p} \gamma^{\alpha} (f - g \gamma_{5}) \psi^{n}.
\label{eq:nucl-current}
\end{equation}
The form factors $f$ and $g$ in Eq.~\eqref{eq:nucl-current}
are taken as constant. In the calculation below,
we shall use the Dirac-Pauli representation and take
the magnetic field $\mathbf{B}$ to be in the positive
\textit{z}-direction. All the terms of order $m_N^{-1}$ and higher are
ignored in the calculation.

The wavefunction of a left-handed neutrino with momentum 
$\mathbf{p}_{\nu} = E_{\nu} (\sin \Theta_{\nu}, 0, \cos \Theta_{\nu})$
is
\begin{equation}
\psi_{\mathbf{p}_{\nu}}^{\nu} =
\frac{e^{i (\mathbf{p}_{\nu} \cdot \mathbf{x}-E_{\nu } t )}}%
{\sqrt{2} L^{3/2}}
\left( 	\begin{array}{c}
 	\sin \frac{\Theta_{\nu}}{2}\\
 	-\cos \frac{\Theta_{\nu}}{2}\\
	 -\sin \frac{\Theta_{\nu}}{2}\\
 	\cos \frac{\Theta_{\nu}}{2}
	\end{array}
\right)
 = \frac{e^{i (\mathbf{p}_{\nu} \cdot \mathbf{x}-E_{\nu } t )}}{L^{3/2}}
U_{\mathbf{p}_{\nu}}^{\nu},
\end{equation}
where $L$ is the linear size of the normalization volume. 
The above wavefunction also applies to a right-handed antineutrino with
the same momentum. The wavefunction of a non-relativistic nucleon is
\begin{equation}
\psi_{\mathbf{k}_{n(p)}, s_{n(p)}}^{n(p)} = 
\frac{e^{i (\mathbf{k}_{n(p)} \cdot \mathbf{x} -m_{n(p)} t)}}{L^{3/2}}
\left(	\begin{array}{c}
 	\delta _{s_{n(p)},1}\\
 	\delta _{s_{n(p)},-1}\\
 	0\\
 	0
	\end{array}
\right) 
= \frac{e^{i (\mathbf{k}_{n(p)} \cdot \mathbf{x} -m_{n(p)} t)}}{L^{3/2}}
U_{\mathbf{k}_{n(p)}, s_{n(p)}}^{n(p)},
\end{equation}
where $s_{n(p)} = \pm 1$ denotes the spin state and $\mathbf{k}_{n(p)}$
is the nucleon momentum. 

In cylindrical coordinates $(\xi, \phi, z)$, 
the wavefunction of an electron is
\begin{equation}
\psi_{n, n_r, p_{ez}, s_{e}}^{e^{-}} = 
\frac{e^{i (p_{ez} z - E_{e} t)} e^{i (n-n_r) \phi}}{\sqrt{2 \pi \lambda^{2}L}}
U_{n, n_r, p_{ez}, s_{e}}^{e^{-}},
\label{eq:electron-wave}
\end{equation}
where $n$ is the quantum number of the Landau level, $n_r$ is the quantum
number of the gyromotion center, $\lambda = 1 / \sqrt{eB}$ is a characteristic
length scale defined by the strength of the magnetic field, and
$s_{e} = \pm 1$ corresponds to the two
spin states when the electron is at rest. The spinor 
$\psi_{n, n_r, p_{ez}, s_{e}}^{e^{-}}$ in Eq.~\eqref{eq:electron-wave} is
\begin{equation}
U_{n, n_r, p_{ez}, s_e=1}^{e^{-}} =
\frac{1}{\sqrt{2 E_e (E_e + m_e)}}
\left(	\begin{array}{c}
 	(m_{e} + E_{e}) e^{-i \phi} I_{n-1,n_r}(\xi^2/2\lambda^2)\\
 	0\\
 	p_{ez} e^{-i \phi} I_{n-1,n_r}(\xi^2/2\lambda^2)\\
 	i \frac{\sqrt{2n}}{\lambda} I_{n,n_r}(\xi^2/2\lambda^2)
	\end{array}
\right)
\end{equation}
and
\begin{equation}
U_{n, n_r, p_{ez}, s_e=-1}^{e^{-}} =
\frac{1}{\sqrt{2 E_e (E_e + m_e)}}
\left(	\begin{array}{c}
 	0\\
 	(m_{e} + E_{e}) I_{n, n_r}(\xi^2/2\lambda^2)\\
 	-i \frac{\sqrt{2n}}{\lambda} e^{-i \phi} 
	I_{n-1, n_r}(\xi^2/2\lambda^2)\\
 	-p_{ez} I_{n, n_r}(\xi^2/2\lambda^2)
	\end{array}
\right).
\end{equation}
The special function $I_{n,n_r}(\eta)$ in the above equations
is defined in Ref.~\cite{Sokolov:1968}, and can
be written in terms of the generalized Laguerre polynomial 
$L_{n_r}^{n-n_r}(\eta)$ as
\begin{equation}
I_{n ,n_r}(\eta) = \sqrt{\frac{n_r!}{n!}} e^{-\eta / 2} \eta^{(n-n_r) / 2}
L_{n_r}^{n-n_r}(\eta).
\end{equation}
The electron wavefunction discussed above is the same as given in
Ref.~\cite{Johnson:1949} up to a phase factor [Note that there are a few
typos in that reference: Eq.~(45) should read 
$\psi_{n,l} = (-i)^{n} (2^{n+l+1} \pi l! n!)^{-\frac{1}{2}}
\lambda^{n-1} \times \cdots$ and Eq.~(46) should read
$\psi_{n,l} = i^{n} (2^{n-l+1} \pi l! n!)^{-\frac{1}{2}}
\lambda^{-1} (\rho / \lambda)^{n-l} \times \cdots$].
The wavefunction of a positron is 
\begin{equation}
\psi_{n, n_r, p_{ez}, s_{e}}^{e^{+}} =
\frac{e^{-i(p_{ez} z - E_{e} t)} e^{i (n-n_r)\phi}}{\sqrt{2 \pi \lambda^{2}L}}
U_{n, n_r, p_{ez}, s_{e}}^{e^{+}},
\end{equation}
where
\begin{equation}
U_{n, n_r, p_{ez}, s_e=1}^{e^{+}} = 
\frac{1}{\sqrt{2 E_e (E_e + m_e)}}
\left(	\begin{array}{c}
 	i \frac{\sqrt{2n}}{\lambda} e^{-i \phi}I_{n-1 ,n_r}(\xi^2/2\lambda^2)\\
 	-p_{ez} I_{n ,n_r}(\xi^2/2\lambda^2)\\
 	0\\
 	(m_{e} + E_{e}) I_{n ,n_r}(\xi^2/2\lambda^2)
	\end{array}
\right),
\end{equation}
and
\begin{equation}
U_{n ,n_r ,p_{ez}, s_e=-1}^{e^{+}} =
\frac{1}{\sqrt{2 E_e (E_e + m_e)}}
\left(	\begin{array}{c}
 	-p_{ez} e^{-i \phi} I_{n-1, n_r}(\xi^2/2\lambda^2)\\
 	i \frac{\sqrt{2n}}{\lambda} I_{n, n_r}(\xi^2/2\lambda^2)\\
 	-(m_{e} + E_{e}) e^{-i \phi} I_{n-1, n_r}(\xi^2/2\lambda^2)\\
 	0
	\end{array}
\right).
\end{equation}
All the wavefunctions are normalized to have one particle in 
a volume of $L^3$.

The scattering matrix of $\nu_{e} + n \rightarrow e^{-} + p$ is
\begin{eqnarray}
i \mathcal{T}_{fi} & = & 
\frac{G_{F} \cos \theta_{C}}{\sqrt{2}}
\int \bar{\psi}_{\mathbf{k}_{p} ,s_{p}}^{p}
\gamma_{\alpha} (f - g \gamma_{5}) \psi_{\mathbf{k}_{n}, s_{n}}^{n}
\bar{\psi}_{n, n_r, p_{ez}, s_{e}}^{e^{-}}
\gamma^{\alpha} (1 - \gamma_{5}) \psi_{\mathbf{p}_{\nu}}^{\nu}
\, \ud^{4} x \nonumber \\
 & = & \frac{G_{F} \cos \theta_{C}}{\sqrt{2}} 
\frac{1}{\sqrt{2 \pi \lambda^{2}} L^{5}} 
2 \pi \delta(E_{e} - E_{\nu} - \Delta) 
2 \pi \delta(k_{pz} + p_{ez} - k_{nz} - p_{\nu z}) \mathfrak{M}.
\label{eq:nun-matrix}
\end{eqnarray}
In Eq.~\eqref{eq:nun-matrix}, 
\begin{equation}
\mathfrak{M} = \int_{0}^{\infty } \xi \,\ud \xi
\int_{0}^{2 \pi} 
e^{i \mathbf{w}_{\perp} \cdot \mathbf{x}_{\perp}} e^{-i (n-n_r) \phi} 
\bar{U}_{\mathbf{k}_{p}, s_{p}}^{p}
\gamma_{\alpha}(f - g \gamma_{5}) U_{\mathbf{k}_{n} ,s_{n}}^{n}
\bar{U}_{n, n_r, p_{ez}, s_{e}}^{e^{-}}
\gamma^{\alpha}(1 - \gamma_{5}) U_{\mathbf{p}_{\nu}}^{\nu}
\, \ud \phi,
\end{equation}
where the subscript ``$\perp$'' denotes a vector in the \textit{xy}-plane and 
$\mathbf{w}_{\perp}=(\mathbf{k}_n+\mathbf{p}_\nu-\mathbf{k}_p)_\perp$. 
Similarly, the scattering matrix of
$e^{+} + n \rightarrow \bar{\nu}_{e} + p$ is,
\begin{eqnarray}
i \mathcal{T}_{fi}^\prime & = & 
\frac{G_{F} \cos \theta_{C}}{\sqrt{2}}
\int \bar{\psi}_{\mathbf{k}_{p}, s_{p}}^{p}
\gamma_{\alpha} (f - g \gamma_{5}) \psi_{\mathbf{k}_{n}, s_{n}}^{n}
\bar{\psi}_{n, n_r, p_{ez}, s_{e}}^{e^{+}}
\gamma^{\alpha} (1 - \gamma_{5}) \psi_{\mathbf{p}_{\nu}}^{\bar{\nu}}
\, \ud^{4} x \nonumber \\
 & = & \frac{G_{F} \cos \theta_{C}}{\sqrt{2}}
\frac{1}{\sqrt{2 \pi \lambda^{2}} L^{5}}
2 \pi \delta (E_{\nu } - E_{e} - \Delta)
2 \pi \delta (k_{pz} + p_{\nu z} - k_{nz} - p_{ez}) \mathfrak{M}^\prime.
\label{eq:en-matrix}
\end{eqnarray}
In Eq.~\eqref{eq:en-matrix},
\begin{equation}
\mathfrak{M}^\prime = 
\int_{0}^{\infty} \xi \,\ud \xi
\int_{0}^{2 \pi }
e^{i \mathbf{w}^\prime_{\perp} \cdot \mathbf{x}_{\perp}} e^{-i (n-n_r) \phi} 
\bar{U}_{\mathbf{k}_{p}, s_{p}}^{p} 
\gamma _{\alpha}(f - g\gamma_{5}) U_{\mathbf{k}_{n}, s_{n}}^{n}
\bar{U}_{n, n_r, p_{ez}, s_{e}}^{e^{+}}
\gamma^{\alpha}(1 - \gamma_{5}) U_{\mathbf{p}_{\nu}}^{\bar{\nu}}
\, \ud \phi ,
\end{equation}
where 
$\mathbf{w}^\prime_{\perp}=(\mathbf{k}_n-\mathbf{k}_p-\mathbf{p}_\nu)_\perp$. 
The scattering matrices of $e^-+p\rightarrow n+\nu_e$ and
$\bar{\nu}_e+p\rightarrow n+e^+$ are the Hermitian conjugate of those
in Eqs.~\eqref{eq:nun-matrix} and \eqref{eq:en-matrix}, respectively.

Based on formula 8.411.1 in Ref.~\cite{Gradshteyn:1980}, we obtain
\begin{eqnarray}
\int_{0}^{2 \pi} 
e^{i \mathbf{w}_{\perp} \cdot \mathbf{x}_{\perp} - i (n-n_r) \phi} 
\, \ud \phi
 & =  & \int_{0}^{2 \pi} 
e^{i w_{\perp} \rho \cos (\phi - \phi_{0}) - i (n-n_r) \phi } 
\,\ud \phi \nonumber \\
& =  & 2 \pi i^{n-n_r} e^{-i (n-n_r) \phi_{0}} J_{n-n_r}(w_{\perp} \xi),
\end{eqnarray}
where $\phi_{0}$ is the azimuthal angle of $\mathbf{w}_{\perp}$,
and $J_n(\eta)$ is the Bessel function of order $n$.
Using this and formula 7.421.4 in Ref.~\cite{Gradshteyn:1980}
\begin{equation}
\int_{0}^{\infty} x^{\nu + 1} e^{-\beta x^{2}} 
L_{n}^{\nu}(\alpha x^{2}) J_{\nu}(x y)\, \ud x = 
2^{-\nu - 1} \beta^{-\nu - n - 1} (\beta - \alpha)^{n} 
y^{\nu} e^{-y^{2} / 4 \beta} L_{n}^{\nu}
\left( \frac{\alpha y^{2}}{4 \beta (\alpha - \beta)} \right),
\end{equation}
one can prove that
\begin{eqnarray}
 &  & \int_{0}^{\infty} \xi \,\ud \xi \int_{0}^{2 \pi} 
e^{i \mathbf{w}_{\perp} \cdot \mathbf{x}_{\perp}}
e^{-i (n-n_r) \phi} I_{n, n_r} (\xi^{2} / 2 \lambda^{2})
\, \ud \phi \nonumber \\
= &  & i^{n + n_r} e^{-i (n-n_r) \phi_{0}} 4\pi \lambda^{2}
I_{n, n_r}(2 \lambda^{2} w_{\perp}^{2}).
\label{eq:int-rel}
\end{eqnarray}
Using Eq.~\eqref{eq:int-rel}, we obtain $\left|\mathfrak{M}\right|^{2}$
of the forward and reverse processes in Eq.~\eqref{eq:nun} as
\begin{subequations}
\begin{eqnarray}
\left| \mathfrak{M} \right|_{s_{p}=1, s_{n}=1}^{2} & = & 
4 (2 \pi \lambda^{2})^{2} \Big [(f+g)^{2} 
\left(1 + v_{ez} \right) (1 + \cos \Theta_{\nu}) 
I_{n, n_r}^2(2 \lambda^{2} w_{\perp}^{2}) \nonumber\\
& & + (f-g)^{2}
\left(1 - v_{ez} \right) (1 - \cos \Theta_{\nu}) 
I_{n-1, n_r}^2(2 \lambda^{2} w_{\perp}^{2}) \nonumber \\
 &  & + 2 (f^{2} - g^{2}) \frac{\sqrt{2neB}}{E_{e}} 
\cos \phi_{0} \sin \Theta_{\nu} I_{n,n_r}(2 \lambda^{2} w_{\perp}^{2}) 
I_{n-1,n_r}(2 \lambda^{2} w_{\perp}^{2}) \Big], \\
\left| \mathfrak{M} \right|_{s_{p}=1, s_{n}=-1}^{2} & = & 
16 (2 \pi \lambda^{2})^{2} g^{2} 
\left( 1 + v_{ez} \right) (1 - \cos \Theta_{\nu}) 
I_{n, n_r}^{2}(2 \lambda^{2} w_{\perp}^{2}),\\
\left| \mathfrak{M} \right|_{s_{p}=-1, s_{n}=1}^{2} & = & 
16 (2 \pi \lambda^{2})^{2} g^{2}
\left( 1 - v_{ez} \right) (1 + \cos \Theta_{\nu}) 
I_{n-1, n_r}^{2}(2 \lambda^{2} w_{\perp}^{2}),\\
\left| \mathfrak{M} \right|_{s_{p}=-1, s_{n}=-1}^{2} & = & 
4 (2 \pi \lambda^{2})^{2} 
\Big [(f-g)^{2} \left( 1 + v_{ez} \right) (1 + \cos \Theta_{\nu}) 
I_{n, n_r}^{2}(2 \lambda^{2} w_{\perp}^{2}) \nonumber \\
 & & + (f+g)^{2} \left( 1 - v_{ez} \right) (1 -\cos \Theta_{\nu}) 
I_{n-1,n_r}^{2}(2 \lambda^{2} w_{\perp}^{2}) \nonumber \\
 &  & + 2 (f^{2} - g^{2}) \frac{\sqrt{2neB}}{E_{e}} 
\cos \phi_{0} \sin \Theta_{\nu} I_{n, n_r}(2 \lambda^{2} w_{\perp}^{2}) 
I_{n-1,n_r}(2 \lambda^{2} w_{\perp}^{2}) \Big],
\end{eqnarray}
\label{eq:nun-amp}
\end{subequations}  
where $v_{ez}=p_{ez}/E_{e}$ is the longitudinal velocity of the electron.
The corresponding expressions of $\left|\mathfrak{M}^\prime\right|^{2}$,
which apply to the forward and reverse processes in Eq.~\eqref{eq:nup}, can
be obtained from the above expressions of $\left|\mathfrak{M}\right|^{2}$
by changing the signs of the terms proportional to $(f^2-g^2)$
and replacing $\mathbf{w}_\perp$ with $\mathbf{w}^\prime_\perp$.

The cross section of $\nu_e+n\rightarrow e^-+p$ is
\begin{eqnarray}
\sigma_{\nu_e n}(B)  & = &
\sum_{s_{p}} \int \frac{L^3 \ud^3 k_{p}}{(2 \pi)^3}
\sum_{n,n_r,s_{e}} \int \frac{L \ud p_{ez}}{2 \pi} 
\frac{1}{L^{-3} L^{-3}} \frac{G_{F}^{2} \cos ^{2} \theta_{C}}{2} \nonumber \\
 & & \times 2\pi \delta (E_e - E_\nu - \Delta) 
2 \pi \delta (k_{pz}+p_{ez}-k_{nz}-p_{\nu z})
\frac{TL}{2 \pi \lambda^{2} L^{10}} 
\frac{\left| \mathfrak{M} \right|^{2}}{T L^{3}}.
\label{eq:nun-xsection-def}
\end{eqnarray}
In this appendix only, the symbol $T$ deontes the duration of a process
such as $\nu_e+n\rightarrow e^- + p$.  Evaluation of $\sigma_{\nu_e n}(B)$
can be simplified using the summation rule \cite{Sokolov:1968}
for the special function $I_{n,n_r}(2\lambda^2w_\perp^2)$ 
in $\left|\mathfrak{M}\right|^{2}$,
\begin{equation}
\sum_{n_r} I_{n, n_r}(\eta) I_{n^{\prime} n_r}(\eta) = \delta_{n, n^{\prime}},
\label{eq:I-sum}
\end{equation}
where $\delta_{n,n^\prime}$ is the Kronecker delta function.
A difficulty in evaluation of $\sigma_{\nu_e n}(B)$
is that the integrand in Eq.~\eqref{eq:nun-xsection-def} is independent of
$(\mathbf{k}_p)_\perp$ in the infinite nucleon mass limit and the integral
diverges. Another difficulty is that there is a remaining factor of $L^{-2}$.
These difficulties arise
because $e^-$ does not have definite transverse canonical momenta and
we drop all the terms of order $m_N^{-1}$ and higher.
However, $e^-$ has definite transverse kinetic momentum squared
\begin{equation}
(\bm{\pi}_e)_\perp^{2} = 
\left(p_{ex} - \frac{y}{2 \lambda ^{2}}\right)^{2} +
\left(p_{ey} + \frac{x}{2 \lambda ^{2}}\right)^{2}
 = (2n+1)eB.
\end{equation}
The limit on $x$ and $y$ then corresponds to a limit of $L/2 \lambda^{2}$
on $p_{ex}$ and $p_{ey}$. Thus, we take
\begin{equation}
\int \frac{L^{2} \ud k_{px}\, \ud k_{py}}{(2 \pi)^{2}}
\longrightarrow 
\frac{L^{4}}{4 \left( 2 \pi \lambda^{2} \right)^{2}}.
\label{eq:int-rplc}
\end{equation}
Using Eqs.~\eqref{eq:nun-amp}, \eqref{eq:nun-xsection-def},
\eqref{eq:I-sum}, and \eqref{eq:int-rplc}, we obtain
\begin{eqnarray}
\sigma_{\nu_e n}(B) & = & 
\sigma_{B}^{(1)} 
\left[ 1 + 2 \chi_n \frac{(f + g) g}{f^{2} + 3 g^{2}} 
\cos \Theta_{\nu } \right] \nonumber \\
 & & + \sigma_{B}^{(2)} 
\left[ \frac{f^{2} - g^{2}}{f^{2} + 3 g^{2}}
\cos \Theta_{\nu} + 2 \chi_n \frac{(f - g)}{f^{2} + 3 g^{2}} \right],
\label{eq:nun-xsection}
\end{eqnarray}
where 
\begin{eqnarray}
\sigma_{B}^{(1)} & = & 
\frac{G_{F}^{2} \cos^{2} \theta_{C}}{2 \pi}(f^{2} + 3 g^{2}) eB
\sum_{n=0}^{n_{\textrm{max}}} 
\frac{g_{n} E_{e}}{\sqrt{E_{e}^{2} - m_{e}^{2} - 2neB}},
\label{eq:sigb1}\\
\sigma_{B}^{(2)} & = & 
\frac{G_{F}^{2} \cos^{2} \theta_{C}}{2 \pi}(f^{2} + 3 g^{2}) eB
\frac{E_{e}}{\sqrt{E_{e}^{2} - m_{e}^{2}}},
\end{eqnarray}
and $\chi_n$ is the net polarization of the neutron in the initial state.
In Eq.~\eqref{eq:sigb1}, $g_n$ denotes the degeneracy 
of Landau level $n$ for $e^-$.
The cross section of $\bar{\nu}_e+p\rightarrow e^+ + n$ can be obtained in a 
similar way as
\begin{eqnarray}
\sigma_{\bar{\nu}_e p}(B) & = & 
\sigma_{B}^{(1)} 
\left[ 1 + 2 \chi_p \frac{(f - g) g}{f^{2} + 3 g^{2}} 
\cos \Theta_{\nu } \right] \nonumber \\
 & & + \sigma_{B}^{(2)} 
\left[ \frac{f^{2} - g^{2}}{f^{2} + 3 g^{2}}
\cos \Theta_{\nu} + 2 \chi_p \frac{(f + g)}{f^{2} + 3 g^{2}} \right].
\label{eq:nup-xsection}
\end{eqnarray}

As $e^-$ does not have definite velocity, we calculate the volume reaction 
rate instead of the cross section for $e^-+p\rightarrow\nu_e+n$.
To illustrate the dependence on the direction of the outgoing $\nu_e$, we
first calculate the  differential volume reaction rate 
\begin{eqnarray}
\frac{\ud \Gamma_{e^- p} (B)}{\ud \! \cos \Theta_{\nu}} & = & 
\overline{\sum_{i}} \sum_{s_n} 
\int \frac{L^{3} \ud^3 k_{n}}{(2 \pi)^{3}} 
\int \frac{L^{3} 2 \pi E_{\nu}^{2} \ud E_{\nu}}{(2 \pi)^{3}}
\frac{1}{L^{-3} L^{-3}}
\frac{G_{F}^{2} \cos^{2} \theta_{C}}{2}
\nonumber \\
 &  & \times 2 \pi \delta (E_\nu + \Delta - E_e) 
2 \pi \delta (p_{\nu z} + k_{nz} - p_{ez} - k_{pz})
\frac{TL}{2 \pi \lambda^{2} L^{10}}
\frac{\left| \mathfrak{M} \right|^{2}}{T L^{3}}.
\label{eq:ep-volrate-def}
\end{eqnarray}
In Eq.~\eqref{eq:ep-volrate-def},
\begin{equation}
\overline{\sum_{i}} = \frac{1}{g_n}\sum_{s_e} \, \frac{1}{2}\sum_{p_{ez}} \,
\frac{2\pi\lambda^2}{L^2}\sum_{n_r},
\label{eq:avg-rplc}
\end{equation}
which represents the average over all possible initial $e^-$ states with
a given energy $E_e$ and a given Landau level quantum number $n$.
Using Eqs.~\eqref{eq:nun-amp}, \eqref{eq:int-rplc} (with $k_{px(y)}$ replaced
with $k_{nx(y)}$), \eqref{eq:ep-volrate-def}, and \eqref{eq:avg-rplc}, 
we obtain
\begin{eqnarray}
\frac{\ud \Gamma_{e^- p} (B)}{\ud \! \cos \Theta_{\nu}} & = & 
\frac{\Gamma_{eN}^{(0)}}{2}
\left[ 1 + 2 \chi_p \frac{(f - g) g}{f^{2} + 3 g^{2}}
\cos \Theta_{\nu } \right] \nonumber \\
 & & + \delta_{n,0} \frac{\Gamma_{eN}^{(0)}}{2}
\left[ \frac{f^{2}-g^{2}}{f^{2} + 3 g^{2}} \cos \Theta_{\nu} +
2 \chi_p \frac{(f + g) g}{f^{2} + 3 g^{2}}\right],
\label{eq:diff-ep-volrate}
\end{eqnarray}
where
\begin{equation}
\Gamma_{eN}^{(0)}=\frac{G_F^2\cos^2\theta_C}{2\pi}(f^2+3g^2)E_\nu^2
\end{equation}
is the volume reaction rate without magnetic field.
Integrating \eqref{eq:diff-ep-volrate} over $\cos \Theta_{\nu }$, we have
\begin{equation}
\Gamma_{e^- p}(B) = \Gamma_{eN}^{(0)}
\left[ 1 + \delta_{n,0} \, 2\chi_p 
\frac{(f + g) g}{f^{2} + 3 g^{2}} \right].
\end{equation}
The volume reaction rate of $e^+ + n \rightarrow \bar{\nu}_e + p$ 
can be  obtained in a similar way as
\begin{eqnarray}
\frac{\ud \Gamma_{e^+ n} (B)}{\ud \! \cos \Theta_{\nu}} & = & 
\frac{\Gamma_{eN}^{(0)}}{2}
\left[ 1 + 2 \chi_n \frac{(f + g) g}{f^{2} + 3 g^{2}}
\cos \Theta_{\nu } \right] \nonumber \\
 & & + \delta_{n,0} \frac{\Gamma_{eN}^{(0)}}{2}
\left[ \frac{f^{2}-g^{2}}{f^{2} + 3 g^{2}} \cos \Theta_{\nu} +
2 \chi_n \frac{(f - g) g}{f^{2} + 3 g^{2}} \right],
\label{eq:diff-en-volrate}
\end{eqnarray}
and
\begin{equation}
\Gamma_{e^+ n}(B) = \Gamma_{eN}^{(0)}
\left[ 1 + \delta_{n,0} \, 2\chi_n
\frac{(f - g) g}{f^{2} + 3 g^{2}} \right].
\end{equation}

\bibliography{ref}

\begin{thebibliography}{27}
\expandafter\ifx\csname natexlab\endcsname\relax\def\natexlab#1{#1}\fi
\expandafter\ifx\csname bibnamefont\endcsname\relax
  \def\bibnamefont#1{#1}\fi
\expandafter\ifx\csname bibfnamefont\endcsname\relax
  \def\bibfnamefont#1{#1}\fi
\expandafter\ifx\csname citenamefont\endcsname\relax
  \def\citenamefont#1{#1}\fi
\expandafter\ifx\csname url\endcsname\relax
  \def\url#1{\texttt{#1}}\fi
\expandafter\ifx\csname urlprefix\endcsname\relax\def\urlprefix{URL }\fi
\providecommand{\bibinfo}[2]{#2}
\providecommand{\eprint}[2][]{\url{#2}}

\bibitem[{\citenamefont{Bethe}(1990)}]{Bethe:1990mw}
\bibinfo{author}{\bibfnamefont{H.~A.} \bibnamefont{Bethe}},
  \bibinfo{journal}{Rev. Mod. Phys.} \textbf{\bibinfo{volume}{62}},
  \bibinfo{pages}{801} (\bibinfo{year}{1990}).

\bibitem[{\citenamefont{Bethe and Wilson}(1985)}]{Bethe:1985}
\bibinfo{author}{\bibfnamefont{H.~A.} \bibnamefont{Bethe}} \bibnamefont{and}
  \bibinfo{author}{\bibfnamefont{J.~R.} \bibnamefont{Wilson}},
  \bibinfo{journal}{Astrophys. J.} \textbf{\bibinfo{volume}{295}},
  \bibinfo{pages}{14} (\bibinfo{year}{1985}).

\bibitem[{\citenamefont{Rampp and Janka}(2000)}]{Rampp:2000}
\bibinfo{author}{\bibfnamefont{M.}~\bibnamefont{Rampp}} \bibnamefont{and}
  \bibinfo{author}{\bibfnamefont{H.-T.} \bibnamefont{Janka}},
  \bibinfo{journal}{Astrophys. J.} \textbf{\bibinfo{volume}{539}},
  \bibinfo{pages}{L33} (\bibinfo{year}{2000}), \eprint{astro-ph/0005438}.

\bibitem[{\citenamefont{Liebendörfer et~al.}(2001)\citenamefont{Liebendörfer,
  Mezzacappa, Thielemann, Messer, Hix, and Bruenn}}]{Liebendorfer:2001}
\bibinfo{author}{\bibfnamefont{M.}~\bibnamefont{Liebendörfer}},
  \bibinfo{author}{\bibfnamefont{A.}~\bibnamefont{Mezzacappa}},
  \bibinfo{author}{\bibfnamefont{F.-K.} \bibnamefont{Thielemann}},
  \bibinfo{author}{\bibfnamefont{O.~E.} \bibnamefont{Messer}},
  \bibinfo{author}{\bibfnamefont{W.~R.} \bibnamefont{Hix}}, \bibnamefont{and}
  \bibinfo{author}{\bibfnamefont{S.~W.} \bibnamefont{Bruenn}},
  \bibinfo{journal}{Phys. Rev. D} \textbf{\bibinfo{volume}{63}},
  \bibinfo{pages}{103004} (\bibinfo{year}{2001}), \eprint{astro-ph/0006418}.

\bibitem[{\citenamefont{Fryer and Warren}(2002)}]{Fryer:2002}
\bibinfo{author}{\bibfnamefont{C.~L.} \bibnamefont{Fryer}} \bibnamefont{and}
  \bibinfo{author}{\bibfnamefont{M.~S.} \bibnamefont{Warren}},
  \bibinfo{journal}{Astrophys. J.} \textbf{\bibinfo{volume}{574}},
  \bibinfo{pages}{L65} (\bibinfo{year}{2002}), \eprint{astro-ph/0206017}.

\bibitem[{\citenamefont{Kouveliotou et~al.}(1999)\citenamefont{Kouveliotou,
  Strohmayer, Hurley, van Paradijs, Finger, Dieters, Woods, Thompson, and
  Duncan}}]{Kouveliotou:1999}
\bibinfo{author}{\bibfnamefont{C.}~\bibnamefont{Kouveliotou}},
  \bibinfo{author}{\bibfnamefont{T.}~\bibnamefont{Strohmayer}},
  \bibinfo{author}{\bibfnamefont{K.}~\bibnamefont{Hurley}},
  \bibinfo{author}{\bibfnamefont{J.}~\bibnamefont{van Paradijs}},
  \bibinfo{author}{\bibfnamefont{M.~H.} \bibnamefont{Finger}},
  \bibinfo{author}{\bibfnamefont{S.}~\bibnamefont{Dieters}},
  \bibinfo{author}{\bibfnamefont{P.}~\bibnamefont{Woods}},
  \bibinfo{author}{\bibfnamefont{C.}~\bibnamefont{Thompson}}, \bibnamefont{and}
  \bibinfo{author}{\bibfnamefont{R.~C.} \bibnamefont{Duncan}},
  \bibinfo{journal}{Astrophys. J.} \textbf{\bibinfo{volume}{510}},
  \bibinfo{pages}{L115} (\bibinfo{year}{1999}), \eprint{astro-ph/9809140}.

\bibitem[{\citenamefont{Gotthelf et~al.}(1999)\citenamefont{Gotthelf, Vasisht,
  and Dotani}}]{Gotthelf:1999}
\bibinfo{author}{\bibfnamefont{E.~V.} \bibnamefont{Gotthelf}},
  \bibinfo{author}{\bibfnamefont{G.}~\bibnamefont{Vasisht}}, \bibnamefont{and}
  \bibinfo{author}{\bibfnamefont{T.}~\bibnamefont{Dotani}},
  \bibinfo{journal}{Astrophys. J.} \textbf{\bibinfo{volume}{522}},
  \bibinfo{pages}{L49} (\bibinfo{year}{1999}), \eprint{astro-ph/9906122}.

\bibitem[{\citenamefont{Ibrahim et~al.}(2003)\citenamefont{Ibrahim, Swank, and
  Parke}}]{Ibrahim:2003}
\bibinfo{author}{\bibfnamefont{A.~I.} \bibnamefont{Ibrahim}},
  \bibinfo{author}{\bibfnamefont{J.~H.} \bibnamefont{Swank}}, \bibnamefont{and}
  \bibinfo{author}{\bibfnamefont{W.}~\bibnamefont{Parke}},
  \bibinfo{journal}{Astrophys. J.} \textbf{\bibinfo{volume}{584}},
  \bibinfo{pages}{L17} (\bibinfo{year}{2003}), \eprint{astro-ph/0210515}.

\bibitem[{\citenamefont{Lai}(2001)}]{Lai:2000at}
\bibinfo{author}{\bibfnamefont{D.}~\bibnamefont{Lai}}, \bibinfo{journal}{Rev.
  Mod. Phys.} \textbf{\bibinfo{volume}{73}}, \bibinfo{pages}{629}
  (\bibinfo{year}{2001}), \eprint{astro-ph/0009333}.

\bibitem[{\citenamefont{Khokhlov et~al.}(1999)\citenamefont{Khokhlov,
  H\"oflich, Oran, Wheeler, Wang, and Chtchelkanova}}]{Khokhlov:1999}
\bibinfo{author}{\bibfnamefont{A.~M.} \bibnamefont{Khokhlov}},
  \bibinfo{author}{\bibfnamefont{P.~A.} \bibnamefont{H\"oflich}},
  \bibinfo{author}{\bibfnamefont{E.~S.} \bibnamefont{Oran}},
  \bibinfo{author}{\bibfnamefont{J.~C.} \bibnamefont{Wheeler}},
  \bibinfo{author}{\bibfnamefont{L.}~\bibnamefont{Wang}}, \bibnamefont{and}
  \bibinfo{author}{\bibfnamefont{A.~Y.} \bibnamefont{Chtchelkanova}},
  \bibinfo{journal}{Astrophys. J.} \textbf{\bibinfo{volume}{524}},
  \bibinfo{pages}{107} (\bibinfo{year}{1999}), \eprint{astro-ph/9904419}.

\bibitem[{\citenamefont{Raffelt}(1996)}]{Raffelt:1996}
\bibinfo{author}{\bibfnamefont{G.~G.} \bibnamefont{Raffelt}},
  \emph{\bibinfo{title}{Stars as Laboratories for Fundamental Physics}}
  (\bibinfo{publisher}{University of Chicago}, \bibinfo{address}{Chicago},
  \bibinfo{year}{1996}).

\bibitem[{\citenamefont{Janka}(1995)}]{Janka:1995}
\bibinfo{author}{\bibfnamefont{H.-T.} \bibnamefont{Janka}},
  \bibinfo{journal}{Astropart. Phys.} \textbf{\bibinfo{volume}{3}},
  \bibinfo{pages}{377} (\bibinfo{year}{1995}), \eprint{astro-ph/9503068}.

\bibitem[{\citenamefont{Janka}(2001)}]{Janka:2001}
\bibinfo{author}{\bibfnamefont{H.-T.} \bibnamefont{Janka}},
  \bibinfo{journal}{Astron. Astrophys.} \textbf{\bibinfo{volume}{368}},
  \bibinfo{pages}{527} (\bibinfo{year}{2001}), \eprint{astro-ph/0008432}.

\bibitem[{\citenamefont{Gvozdev and Ognev}(2002)}]{Gvozdev:2002}
\bibinfo{author}{\bibfnamefont{A.~A.} \bibnamefont{Gvozdev}} \bibnamefont{and}
  \bibinfo{author}{\bibfnamefont{I.~S.} \bibnamefont{Ognev}},
  \bibinfo{journal}{JETP} \textbf{\bibinfo{volume}{94}}, \bibinfo{pages}{1043}
  (\bibinfo{year}{2002}), \eprint{astro-ph/0403011}.

\bibitem[{\citenamefont{Landau and Lifshitz}(1977)}]{Landau:1977}
\bibinfo{author}{\bibfnamefont{L.~D.} \bibnamefont{Landau}} \bibnamefont{and}
  \bibinfo{author}{\bibfnamefont{E.~M.} \bibnamefont{Lifshitz}},
  \emph{\bibinfo{title}{Quantum Mechanics: non-relativistic theory}}
  (\bibinfo{publisher}{Pergamon}, \bibinfo{address}{Oxford},
  \bibinfo{year}{1977}), \bibinfo{edition}{3rd} ed.

\bibitem[{\citenamefont{Johnson and Lippmann}(1949)}]{Johnson:1949}
\bibinfo{author}{\bibfnamefont{M.~H.} \bibnamefont{Johnson}} \bibnamefont{and}
  \bibinfo{author}{\bibfnamefont{B.~A.} \bibnamefont{Lippmann}},
  \bibinfo{journal}{Phys. Rev.} \textbf{\bibinfo{volume}{76}},
  \bibinfo{pages}{828} (\bibinfo{year}{1949}).

\bibitem[{\citenamefont{Duan and Qian}(2004)}]{dq}
\bibinfo{author}{\bibfnamefont{H.}~\bibnamefont{Duan}} \bibnamefont{and}
  \bibinfo{author}{\bibfnamefont{Y.-Z.} \bibnamefont{Qian}}
  (\bibinfo{year}{2004}), \bibinfo{note}{to be submitted to Phys. Rev. D}.

\bibitem[{\citenamefont{Cordes and Chernoff}(1998)}]{Cordes:1998}
\bibinfo{author}{\bibfnamefont{J.~M.} \bibnamefont{Cordes}} \bibnamefont{and}
  \bibinfo{author}{\bibfnamefont{D.~F.} \bibnamefont{Chernoff}},
  \bibinfo{journal}{Astrophys. J.} \textbf{\bibinfo{volume}{505}},
  \bibinfo{pages}{315} (\bibinfo{year}{1998}), \eprint{astro-ph/9707308}.

\bibitem[{\citenamefont{Chandra et~al.}(2002)\citenamefont{Chandra, Goyal, and
  Goswami}}]{Chandra:2002}
\bibinfo{author}{\bibfnamefont{D.}~\bibnamefont{Chandra}},
  \bibinfo{author}{\bibfnamefont{A.}~\bibnamefont{Goyal}}, \bibnamefont{and}
  \bibinfo{author}{\bibfnamefont{K.}~\bibnamefont{Goswami}},
  \bibinfo{journal}{Phys. Rev. D} \textbf{\bibinfo{volume}{65}},
  \bibinfo{pages}{053003} (\bibinfo{year}{2002}), \eprint{hep-ph/0109057}.

\bibitem[{\citenamefont{Leinson and P\'erez}(1998)}]{Leinson:1998}
\bibinfo{author}{\bibfnamefont{L.~B.} \bibnamefont{Leinson}} \bibnamefont{and}
  \bibinfo{author}{\bibfnamefont{A.}~\bibnamefont{P\'erez}},
  \bibinfo{journal}{JHEP} \textbf{\bibinfo{volume}{9809}}, \bibinfo{pages}{020}
  (\bibinfo{year}{1998}), \eprint{astro-ph/9711216}.

\bibitem[{\citenamefont{Gvozdev and Ognev}(1999)}]{Gvozdev:1999}
\bibinfo{author}{\bibfnamefont{A.~A.} \bibnamefont{Gvozdev}} \bibnamefont{and}
  \bibinfo{author}{\bibfnamefont{I.~S.} \bibnamefont{Ognev}},
  \bibinfo{journal}{JETP Lett} \textbf{\bibinfo{volume}{69}},
  \bibinfo{pages}{365} (\bibinfo{year}{1999}), \eprint{astro-ph/9909154}.

\bibitem[{\citenamefont{Roulet}(1998)}]{Roulet:1998sw}
\bibinfo{author}{\bibfnamefont{E.}~\bibnamefont{Roulet}},
  \bibinfo{journal}{JHEP} \textbf{\bibinfo{volume}{01}}, \bibinfo{pages}{013}
  (\bibinfo{year}{1998}), \eprint{hep-ph/9711206}.

\bibitem[{\citenamefont{Lai and Qian}(1998)}]{Lai:1998sz}
\bibinfo{author}{\bibfnamefont{D.}~\bibnamefont{Lai}} \bibnamefont{and}
  \bibinfo{author}{\bibfnamefont{Y.-Z.} \bibnamefont{Qian}},
  \bibinfo{journal}{Astrophys. J.} \textbf{\bibinfo{volume}{505}},
  \bibinfo{pages}{844} (\bibinfo{year}{1998}), \eprint{astro-ph/9802345}.

\bibitem[{\citenamefont{Arras and Lai}(1999)}]{Arras:1998mv}
\bibinfo{author}{\bibfnamefont{P.}~\bibnamefont{Arras}} \bibnamefont{and}
  \bibinfo{author}{\bibfnamefont{D.}~\bibnamefont{Lai}},
  \bibinfo{journal}{Phys. Rev. D} \textbf{\bibinfo{volume}{60}},
  \bibinfo{pages}{043001} (\bibinfo{year}{1999}), \eprint{astro-ph/9811371}.

\bibitem[{\citenamefont{Bhattacharya and Pal}(2003)}]{Bhattacharya:2002qf}
\bibinfo{author}{\bibfnamefont{K.}~\bibnamefont{Bhattacharya}}
  \bibnamefont{and} \bibinfo{author}{\bibfnamefont{P.~B.} \bibnamefont{Pal}}
  (\bibinfo{year}{2003}), \eprint{hep-ph/0209053}.

\bibitem[{\citenamefont{Sokolov and Temov}(1968)}]{Sokolov:1968}
\bibinfo{author}{\bibfnamefont{A.~A.} \bibnamefont{Sokolov}} \bibnamefont{and}
  \bibinfo{author}{\bibfnamefont{I.~M.} \bibnamefont{Temov}},
  \emph{\bibinfo{title}{Synchrontron Radiation}} (\bibinfo{publisher}{Pergamon,
  Oxford}, \bibinfo{year}{1968}).

\bibitem[{\citenamefont{Gradshteyn and Ryzhik}(1980)}]{Gradshteyn:1980}
\bibinfo{author}{\bibfnamefont{I.~S.} \bibnamefont{Gradshteyn}}
  \bibnamefont{and} \bibinfo{author}{\bibfnamefont{I.~M.}
  \bibnamefont{Ryzhik}}, \emph{\bibinfo{title}{Table of integrals, seriers and
  products}} (\bibinfo{publisher}{Academic}, \bibinfo{address}{New York},
  \bibinfo{year}{1980}).

\end{thebibliography}

\end{document}